\documentclass[twocolumn]{aastex6}
\graphicspath{{/media/alberto/TOSHIBA2/UDS/},{/media/alberto/TRANSCEND/UDS/figures/}}
\usepackage{comment}
\usepackage{multirow}
\usepackage{array}
\usepackage{afterpage}

\newcommand\xmm{XMM$-$\textit{Newton}}
\newcommand\chandra{\textit{Chandra}}

\shorttitle{The NuSTAR survey of UDS field: catalog and CT fraction}
\shortauthors{Masini et al.}
\begin{document}

\title{The NuSTAR Extragalactic Surveys: source catalog and  the \\ Compton-thick fraction in the UDS field}
\author{A. Masini\altaffilmark{1,2,3}, F.~Civano\altaffilmark{3,4}, A.~Comastri\altaffilmark{1}, F.~Fornasini\altaffilmark{3}, D.~R.~Ballantyne\altaffilmark{5}, G.~B.~Lansbury\altaffilmark{6}, E.~Treister\altaffilmark{7}, D.~M.~Alexander\altaffilmark{8}, P.~G.~Boorman\altaffilmark{9}, W.~N.~Brandt\altaffilmark{10,11,12}, D.~Farrah\altaffilmark{13}, P.~Gandhi\altaffilmark{9}, F.~A.~Harrison\altaffilmark{14}, R.~C.~Hickox\altaffilmark{15}, D.~D.~Kocevski\altaffilmark{16}, L.~Lanz\altaffilmark{15}, S.~Marchesi\altaffilmark{17}, S.~Puccetti\altaffilmark{18}, C.~Ricci\altaffilmark{7,19,20}, C.~Saez\altaffilmark{21}, D.~Stern\altaffilmark{22}, and L.~Zappacosta\altaffilmark{23}}

\altaffiltext{1}{INAF$-$Osservatorio Astronomico di Bologna, via Gobetti 93/3, 40129 Bologna, Italy}
\altaffiltext{2}{Dipartimento di Fisica e Astronomia (DIFA), Universit\`a  di Bologna,  via Gobetti 93/2, 40129 Bologna, Italy}
\altaffiltext{3}{Harvard$-$Smithsonian  Center  for  Astrophysics,  60 Garden Street, Cambridge, MA 02138, USA}
\altaffiltext{4}{Yale Center for Astronomy and Astrophysics, 260 Whitney Avenue, New Haven, CT 06520, USA}
\altaffiltext{5}{Center for Relativistic Astrophysics, School of Physics, Georgia Institute of Technology, 837 State Street, Atlanta, GA 30332$-$0430, USA}
\altaffiltext{6}{Institute of Astronomy, University of Cambridge, Madingley Road, Cambridge, CB3 0HA, UK}
\altaffiltext{7}{Instituto de Astrof\'isica, Facultad de F\'isica, Pontificia Universidad Cat\'olica de Chile, Casilla 306, Santiago 22, Chile}
\altaffiltext{8}{Centre for Extragalactic Astronomy, Department of Physics, Durham University, South Road, Durham, DH1 3LE, UK}
\altaffiltext{9}{Department of Physics \& Astronomy, Faculty of Physical Sciences and Engineering, University of Southampton, Southampton, SO17 1BJ, UK}
\altaffiltext{10}{Department of Astronomy and Astrophysics, 525 Davey Lab, The Pennsylvania State University, University Park, PA
16802, USA}
\altaffiltext{11}{Institute for Gravitation and the Cosmos, The Pennsylvania State University, University Park, PA 16802, USA}
\altaffiltext{12}{Department of Physics, 104 Davey Laboratory, The Pennsylvania State University, University Park, PA 16802, USA}
\altaffiltext{13}{Department of Physics,  Virginia Tech,  Blacksburg,  VA 24061, USA}
\altaffiltext{14}{Cahill Center for Astrophysics, 1216 East California Boulevard, California Institute of Technology, Pasadena, CA 91125, USA}
\altaffiltext{15}{Department of Physics and Astronomy, Dartmouth College, 6127 Wilder Laboratory, Hanover, NH 03755, USA}
\altaffiltext{16}{Department of Physics and Astronomy, Colby College, Waterville, ME 04961, USA}
\altaffiltext{17}{Department of Physics \& Astronomy, Clemson University, Clemson, SC 29634, USA}
\altaffiltext{18}{Agenzia Spaziale Italiana$-$Unit\`a di Ricerca Scientifica, Via del Politecnico, 00133 Roma, Italy}
\altaffiltext{19}{EMBIGGEN Anillo, Concepcion, Chile}
\altaffiltext{20}{Kavli Institute for Astronomy and Astrophysics, Peking University, Beijing 100871, China}
\altaffiltext{21}{Observatorio Astron\'omico Cerro Cal\'an, Departamento de Astronom\'ia, Universidad de Chile, Casilla 36$-$D, Santiago, Chile}
\altaffiltext{22}{Jet Propulsion Laboratory, California Institute of Technology, 4800 Oak Grove Drive, Mail Stop 169-221, Pasadena, CA 91109, USA}
\altaffiltext{23}{INAF$-$Osservatorio Astronomico di Roma, via Frascati 33, 00078 Monte Porzio Catone (RM), Italy}
\begin{abstract}

We present the results and the source catalog of the NuSTAR survey in the UKIDSS Ultra Deep Survey (UDS) field, bridging the gap in depth and area between NuSTAR's ECDFS and COSMOS surveys. The survey covers a $\sim 0.6$ deg$^2$ area of the field for a total observing time of $\sim$ 1.75 Ms, to a half-area depth of $\sim$ 155 ks corrected for vignetting at $3-24$ keV, and reaching sensitivity limits at half-area in the full ($3-24$ keV), soft ($3-8$ keV) and hard ($8-24$ keV) bands of $2.2 \times 10^{-14}$ erg cm$^{-2}$  s$^{-1}$, $1.0 \times 10^{-14}$ erg cm$^{-2}$ s$^{-1}$, and $2.7 \times 10^{-14}$ erg cm$^{-2}$ s$^{-1}$, respectively. A total of 67 sources are detected in at least one of the three bands, 56 of which have a robust optical redshift with a median of $\langle z\rangle \sim 1.1$. Through a broadband ($0.5-24$ keV) spectral analysis of the whole sample combined with the NuSTAR hardness ratios, we compute the observed Compton-thick (CT; $N_{\rm H} > 10^{24}$ cm$^{-2}$) fraction. Taking into account the uncertainties on each $N_{\rm H}$ measurement, the final number of CT sources is $6.8\pm1.2$. This corresponds to an observed CT fraction of $(11.5\pm2.0)\%$, providing a robust lower limit to the intrinsic fraction of CT AGN and placing constraints on cosmic X-ray background synthesis models.

\end{abstract}

\section{Introduction} \label{sec:intro}

Supermassive black holes (SMBHs) accreting matter in the centers of galaxies radiate across the electromagnetic spectrum as active galactic nuclei (AGN). Due to a combination of their high luminosities at X-ray wavelengths ($L_{\rm X} \gtrsim 10^{42}$ erg s$^{-1}$) and low dilution from their host galaxies, AGN are effectively detected, traced and studied by X-ray surveys \citep{brandtalexander15}. Indeed, in the past decades, the advent of \xmm\ and \chandra\ was a breakthrough in AGN research, and dozens of X-ray surveys covered a wide range in the flux-area plane \citep[see Figure 16 of][]{civano16}, exploring a large range in redshift and luminosity. These works allowed the luminosity function of AGN to be measured up to $z \sim 5$, both for unobscured ($N_{\rm H} < 10^{22}$ cm$^{-2}$) and obscured ($N_{\rm H} > 10^{22}$ cm$^{-2}$) sources \citep[e.g.,][]{ueda03,lafranca05,polletta06,vito14,marchesi16}. \par However, these surveys were biased against the detection of AGN obscured by large columns of gas ($N_{\rm H} > 10^{24}$ cm$^{-2}$), called Compton-thick (CT) AGN, mainly in the local Universe, up to $z \lesssim 1$. This class of heavily obscured AGN is difficult to study due to the heavy suppression of the spectrum \citep[see, e.g.,][]{teng15}, but it plays a crucial role both in evolutionary models \citep{sanders88,alexanderhickox12,ricci17} and in population synthesis models aiming to explain the shape and intensity of the Cosmic X-ray Background \citep[CXB,][]{gilli07}, of which AGN are the major contributors \citep{comastri95}. Indeed, the integrated emission of a large population of CT AGN would produce a bulk reflection spectrum with a characteristic peak at $20-30$ keV in the overall CXB spectrum, consistently reproducing its peak of emission \citep{madau94,gilli01,treisterurry05}. \par The fraction of CT AGN is then a key observable for X-ray surveys, but its determination suffers degeneracies and observational biases. In particular, high observed energies ($>$ 10 keV) are required to detect CT AGN, at least for redshift $z < 1$. Such X-ray surveys with non-focusing X-ray observatories (e.g., \textit{Swift}/BAT and INTEGRAL) have detected a sizable number of CT AGN \citep{burlon11,ricci15}. However, they directly resolved only a small fraction ($\sim 1-2\%$) of the CXB peak into individual AGN \citep{krivonos07,ajello08,bottacini12}, the majority of which lie in the local Universe ($z < 0.1$), which may not be representative of the whole population.  \par
NuSTAR is the first focusing hard X-ray telescope in orbit, and is composed of two focal plane modules (FPM), referred to as FPMA and FPMB. With the advent of NuSTAR \citep{harrison13}, sensitive hard X-ray surveys above 10 keV started to be feasible, and allowed pushing the search for CT AGN beyond the local Universe, directly resolving 35\% of the CXB in the $8-24$ keV band \citep{harrison16}. \par A wedding-cake strategy for the NuSTAR surveys was adopted: a shallow, wide area survey of the Cosmic Evolution Survey field \citep[COSMOS,][C15 hereafter]{civano15}, a deep, pencil-beam survey of the Extended \chandra\ Deep Field-South \citep[ECDFS,][]{mullaney15}, and a Serendipitous survey \citep{alexander13,lansbury17} were the first steps of a comprehensive survey program, which is now complemented by the observations of the Extended Groth Strip (EGS, Aird et al. in prep), \chandra\ Deep Field-North (CDFN, Del Moro et al. in prep) and Ultra Deep Survey (UDS) fields.
\par
\begin{figure}
\plotone{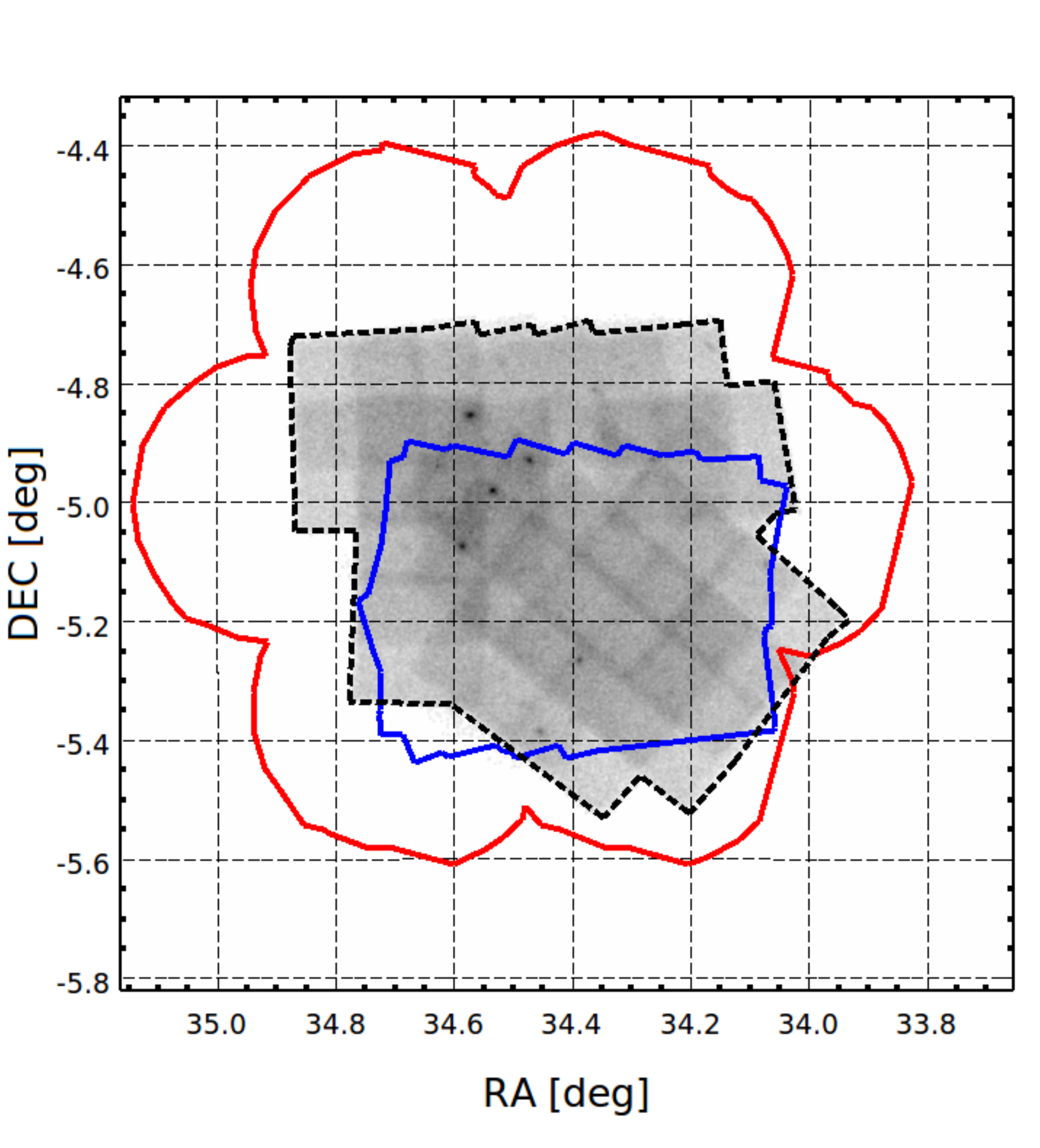}
\caption{Coverage of the UDS field in the X-ray band. The NuSTAR coverage (dashed black) is compared with the flower-shaped \xmm\ (red) and the \chandra\ (blue) coverages. \label{fig:uds_grid}}
\end{figure}

 In this paper, we report on the NuSTAR survey of the UDS field. This field is the deepest component of the UKIRT Infrared Deep Sky Survey \citep[UKIDSS;][]{lawrence07,almaini07}, and has an extensive multi-wavelength coverage. In the radio band, there is Very Large Array (VLA) coverage at 1.4 GHz \citep{simpson06}. Submillimeter coverage comes from the SCUBA Half-Degree Extragalactic Survey (SHADES) survey of the central region of the UDS field \citep{coppin06}. The infrared (IR) band is the most covered, with both ground-based and in-orbit facilities: \textit{Herschel} observed the UDS field as part of the HerMES program \citep{oliver12}, while \textit{Spitzer} observed UDS within the SWIRE survey \citep{lonsdale03} and, more recently, within the SpUDS Spitzer Legacy Survey (PI: Dunlop). Ground-based IR facilities observed the field, primarily the UKIRT WFCAM \citep{casali07} and VISTA, as part of the VIDEO survey \citep{jarvis13}. Also \textit{Hubble} Space Telescope (HST) WFC3 coverage is available \citep{galametz13}, together with deep optical Subaru Suprime-cam imaging data \citep{furusawa08}. Coverage in the $U-$band is provided by the CFHT Megacam (PIs: Almaini, Foucaud). In the X-ray band, the NuSTAR coverage is centered on the coordinates (J2000) R.A. $= 34.4$ deg and DEC. $= -5.1$ deg, and it overlaps with the Subaru \xmm\ Deep Survey \citep[SXDS,][]{ueda08} and \chandra\ UDS survey (Kocevski et al. 2017 submitted) fields. The different X-ray coverages are shown in Figure \ref{fig:uds_grid}.
\par We assume a flat $\Lambda$CDM cosmology ($H_0 = 70$ km s$^{-1}$ Mpc$^{-1}$, $\Omega_{\rm M} = 0.3$, $\Omega_{\Lambda} = 0.7$) throughout the paper, which is organized as follows. In \S \ref{sec:datareduction} the data reduction of all the observations making up the survey is presented. Section \ref{sec:simul} presents the simulations performed to explore the detection parameter space; the results of the source detection in the UDS field are presented in \S \ref{sec:detection}, and compared with \xmm\ and \chandra\ catalogs in \S \ref{sec:match}. The obscuration properties of the sample are presented in \S \ref{sec:obsfrac}, while the measured CT fraction is discussed in \S \ref{sec:discussion}. 
Final remarks are given in \S \ref{sec:conclusions}, while the Appendix shows the catalog description (Appendix \ref{sec:catalog}). Uncertainties are quoted at $1\sigma$ confidence level throughout the paper, unless otherwise stated (e.g., when referring to spectral analysis results).

\section{Data reduction} \label{sec:datareduction}

The NuSTAR UDS survey consists of 35 observations, completed during two different passes, tiled with a half-Field of View (FoV) shift strategy to provide a relatively uniform coverage, despite the roll angle changed significantly between the two passes. The first pass on the field (20 pointings) was performed between January-February 2016, while the second pass (15 pointings) was completed in October-November 2016, for a total observing time of $\sim$ 1.75 Ms. A summary of the observations is presented in Table \ref{tab:obsids}.

\begin{deluxetable}{lccccc}
\tabletypesize{\scriptsize}
\tablecaption{Details of the individual UDS observations. OBSIDs marked with an asterisk are those affected by high background flares.\label{tab:obsids}}
\tablehead{
\colhead{OBSID} & \colhead{Date} & \colhead{R.A. } & \colhead{DEC. } & \colhead{Roll angle } & \colhead{$t_{\rm exp}$ } \\
\colhead{} & \colhead{} & \colhead{[deg]} & \colhead{[deg]} & \colhead{[deg]} & \colhead{[ks]}
}
\colnumbers
\startdata
60111001002* & 2016-01-24 & 34.0911 & $-5.1991$ & 319.3 & 45.8 \\
60111002002 & 2016-01-25 & 34.1591 & $-5.2775$ & 319.3 & 49.9 \\
60111003002 & 2016-01-28 & 34.1613 & $-5.137$  & 319.3  & 49.5 \\
60111004001 & 2016-01-29 & 34.1838 & $-4.9979$ & 319.3 & 50.1 \\
60111005002 & 2016-02-03 & 34.2129 & $-5.3638$ & 319.2 & 51.6 \\
60111006001* & 2016-02-04 & 34.2336 & $-5.2305$ & 319.1 & 49.8 \\
60111007002 & 2016-02-06 & 34.2434 & $-5.0723$ & 319.2 & 48.2 \\
60111008001* & 2016-02-07 & 34.2905 & $-5.2937$ & 319.3 & 45.4 \\
60111009002 & 2016-02-08 & 34.3116 & $-5.1499$ & 319.3 & 48.1 \\
60111010001 & 2016-02-09 & 34.3165 & $-5.0071$ & 319.2 & 48.0 \\
60111011002* & 2016-02-11 & 34.3548 & $-5.3761$ & 319.2 & 46.5 \\
60111012001* & 2016-02-13 & 34.3691 & $-5.2269$ & 319.2  & 43.3 \\
60111013002* & 2016-02-16 & 34.384  & $-5.0892$ & 319.3 & 44.3 \\
60111014001* & 2016-02-17 & 34.3973 & $-4.9459$ & 319.3 & 43.8 \\
60111015002* & 2016-02-21 & 34.44   & $-5.3067$ & 324.2 & 51.0 \\
60111016001* & 2016-02-22 & 34.4424 & $-5.1637$ & 324.2 & 55.7 \\
60111017002 & 2016-02-24 & 34.4632 & $-5.0225$ & 324.3 & 51.0 \\
60111018001 & 2016-02-26 & 34.5089 & $-5.2441$ & 324.3 & 49.8 \\
60111019002* & 2016-02-27 & 34.5231 & $-5.0892$ & 324.3 & 48.9 \\
60111020001 & 2016-02-29 & 34.5345 & $-4.9615$ & 324.3 & 50.5 \\
60111031002* & 2016-10-01 & 34.6363 & $-5.2345$ & 175.5 & 51.9 \\
60111032002* & 2016-10-02 & 34.6439 & $-5.1313$ & 177.5 & 48.4 \\
60111033002 & 2016-10-03 & 34.6426 & $-5.0273$ & 179.5 & 48.8 \\
60111034002 & 2016-10-05 & 34.5369 & $-5.0254$ & 179.4 & 50.5 \\
60111035001* & 2016-10-06 & 34.5345 & $-4.9285$ & 179.4 & 50.1 \\
60111036002 & 2016-10-08 & 34.6373 & $-4.9261$ & 180.5  & 51.2 \\
60111037001 & 2016-10-09 & 34.7387 & $-4.9268$ & 182.5 & 50.3 \\
60111038001 & 2016-10-10 & 34.7404 & $-4.8277$ & 184.4 & 51.2 \\
60111039001* & 2016-10-12 & 34.6419 & $-4.8248$ & 187.4 & 50.7 \\
60111040001* & 2016-10-13 & 34.5355 & $-4.8189$ & 190.3 & 49.9 \\
60111041001 & 2016-10-14 & 34.4367 & $-4.8301$ & 194.4 & 51.2 \\
60111042002* & 2016-11-14 & 34.3471 & $-4.804$  & 273.9 & 50.0 \\
60111043001 & 2016-11-15 & 34.249  & $-4.8016$ & 275.9 & 51.3 \\
60111044001 & 2016-11-17 & 34.2525 & $-4.9036$ & 277.4 & 51.7 \\
60111045001 & 2016-11-18 & 34.1518 & $-4.9055$ & 278.9 & 52.1 \\
\enddata
\tablecomments{\newline (1) Observation ID. \newline (2) Observation's start date. \newline (3)$-$(5) Coordinates and roll angle for each pointing. \newline (6) Exposure time for FPMA, corrected for flaring episodes.}
\end{deluxetable}

\subsection{Flaring episodes} \label{subsec:flares}

The raw event files are processed using the \texttt{nupipeline} task available in the NuSTAR Data Analysis Software (NuSTARDAS\footnote{\url{https://heasarc.gsfc.nasa.gov/docs/nustar/analysis/nustar_swguide.pdf}}). Following C15, full-field lightcurves in the $3.5-9.5$ keV energy band with a binsize of 500 s are produced in order to look for high-background time intervals. Sixteen OBSIDs result affected by background flares (i.e., with a count rate more than a factor of $\sim$ 2 higher than the average, quiescent state) after the analysis of the lightcurves; they are labeled with an asterisk in Table \ref{tab:obsids}.  After cleaning for Good Time Intervals (GTI), the time loss is 39 ks for each focal plane module (FPM), 2.2\% of the total time, resulting in a total cleaned observing time of 1.730 Ms and 1.726 Ms for FPMA and FPMB, respectively.

\subsection{Data, exposure, and background mosaics} \label{subsec:mosaics}

After cleaning for flaring background episodes, we run again \texttt{nupipeline} taking into account the GTI in order to have the final list of cleaned event files. For each observation, we produce images in the $3-24$ keV, $3-8$ keV, $8-24$ keV, $8-16$ keV, $16-24$ keV, and $35-55$ keV energy bands. We will refer to these bands as full (F), soft (S), hard (H), hard-one (H1), hard-two (H2), and very-hard (VH) bands, respectively. The motivation in splitting the H band into two sub-bands comes from multiple sides. On one side, the background contribution is limited in the H1 band, allowing some sources to be more significantly detected narrowing the band; on the other hand, selecting sources at $\sim 15-20$ keV in the H2 band helps us selecting directly those AGN contributing the most to the peak of the CXB. Given their importance, it is worth exploring the feasibility of detecting them with NuSTAR, despite such sources being difficult to be firmly detected, as will be discussed later on (see \S\ref{sec:detection}). \par
\begin{figure}
\plotone{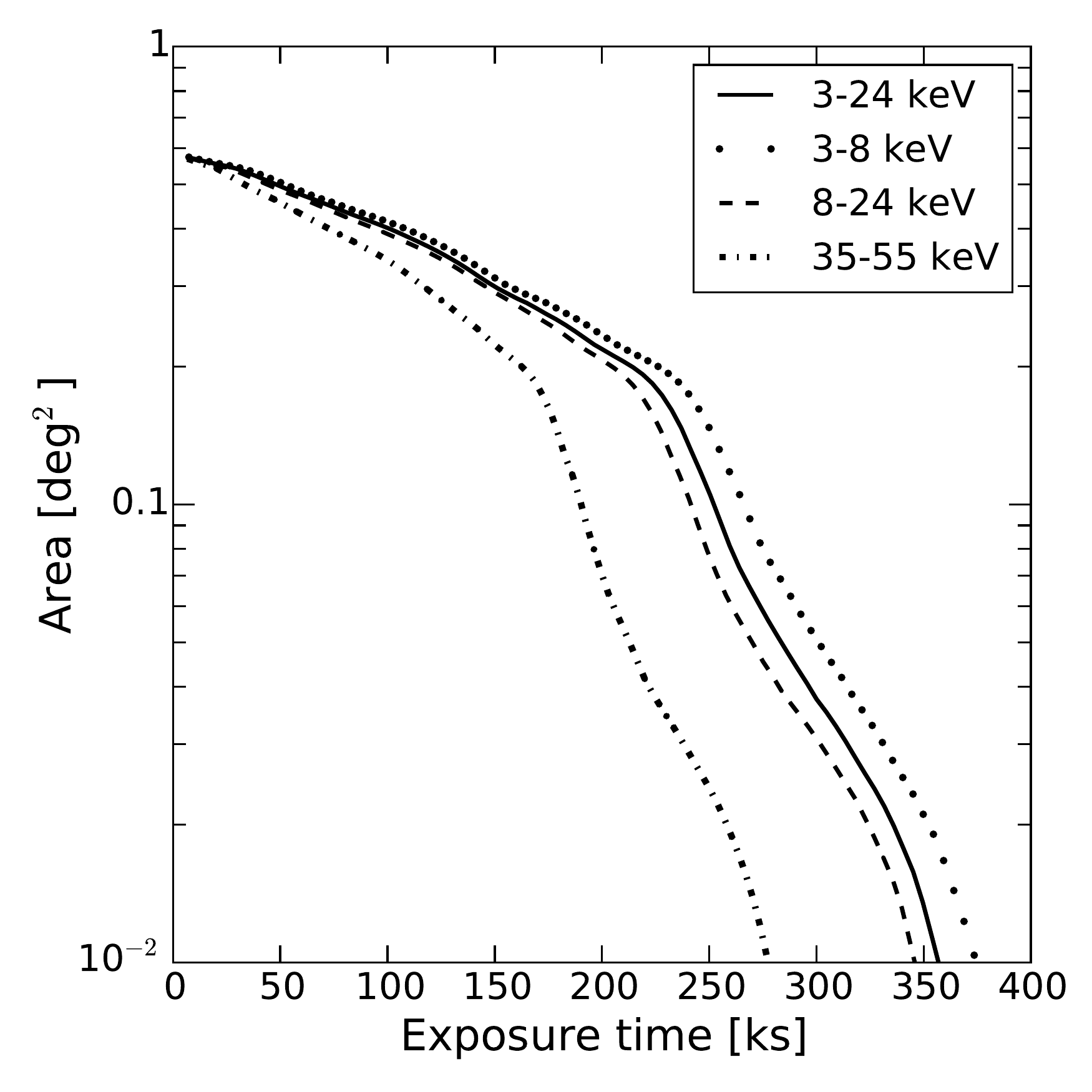}
\caption{Cumulative survey area as a function of exposure depth (FPMA+FPMB), for different energy bands. The total area is 0.58 deg$^2$, and the depth at half area is $\sim 155$ ks in the full ($3-24$ keV) band.}
\label{fig:areavsexpoo}
\end{figure}

 Since the effective area is a continuous function of energy, and producing an exposure map at every energy is extremely time-consuming, we weight the exposure map in every band with an average energy, obtained by convolving the NuSTAR instrumental response with a power law of photon index $\Gamma=1.8$ \citep[i.e., the typical photon index value measured in local AGN;\footnote{It is still unclear if Compton-thick AGN follow the same $\Gamma$ distribution of less obscured ones. NuSTAR observations of a small sample of local Compton-thick AGN present an average $\bar{\Gamma}=1.95$ with a dispersion of $\sigma=0.25$; considering only those sources with a constrained column density, these values become $\bar{\Gamma}=1.89$ and $\sigma=0.19$ \citep{masini16,brightman16}. We therefore assume that $\Gamma=1.8$ is suitable to represent the whole population of AGN.} see][]{burlon11}. Exposure maps are created in the F, S, H, and VH energy bands with the \texttt{nuexpomap} task. Adopting the H exposure map for the H1 and H2 bands results in an underestimation of the exposure of at most 3\%, and overestimation of the exposure of at most 12\%, respectively. A plot of the survey area as a function of vignetting-corrected exposure time is shown in Figure \ref{fig:areavsexpoo}.

Following the general strategy adopted for all the contiguous NuSTAR surveys, we use the \texttt{nuskybgd} software \citep{wik14} to model the background in each energy band. As explained in \citet{wik14}, the NuSTAR background is the sum of different components: below 20~keV, its signal is dominated by photons which are not focused by the mirrors and leak through the open structure of the telescope producing a spatially-dependent pattern (i.e., aperture background). There are also solar photons, a neutron background, and a minor contribution from the focused, but unresolved, sources of the CXB (i.e., fCXB). Above $\sim$ 20~keV, the background is predominantly instrumental, and is composed of a nearly flat power law ($\Gamma \approx 0$) with a forest of activation lines, most notably between $\sim25-35$ keV. This is the reason why the usual NuSTAR surveys are performed in the $3-24$ keV band. We further decide to explore the energy range between the end of the strong instrumental lines, at 35 keV, up to 55 keV, where the NuSTAR effective area starts to decrease substantially. \par We extract background spectra from four $160\arcsec-$radius circular regions, one for each quadrant, avoiding chip gaps. Once the user-defined regions are provided, the \texttt{nuskybgd} software extracts and fits their spectra in XSPEC \citep[v 12.9.1,][]{arnaud96} with the appropriate model, and saves the best-fit parameters. These parameters can be used to extrapolate and produce a background spectrum in a particular region of the FoV, or to produce a background image of the entire FoV.
\par
\begin{figure*}
\plotone{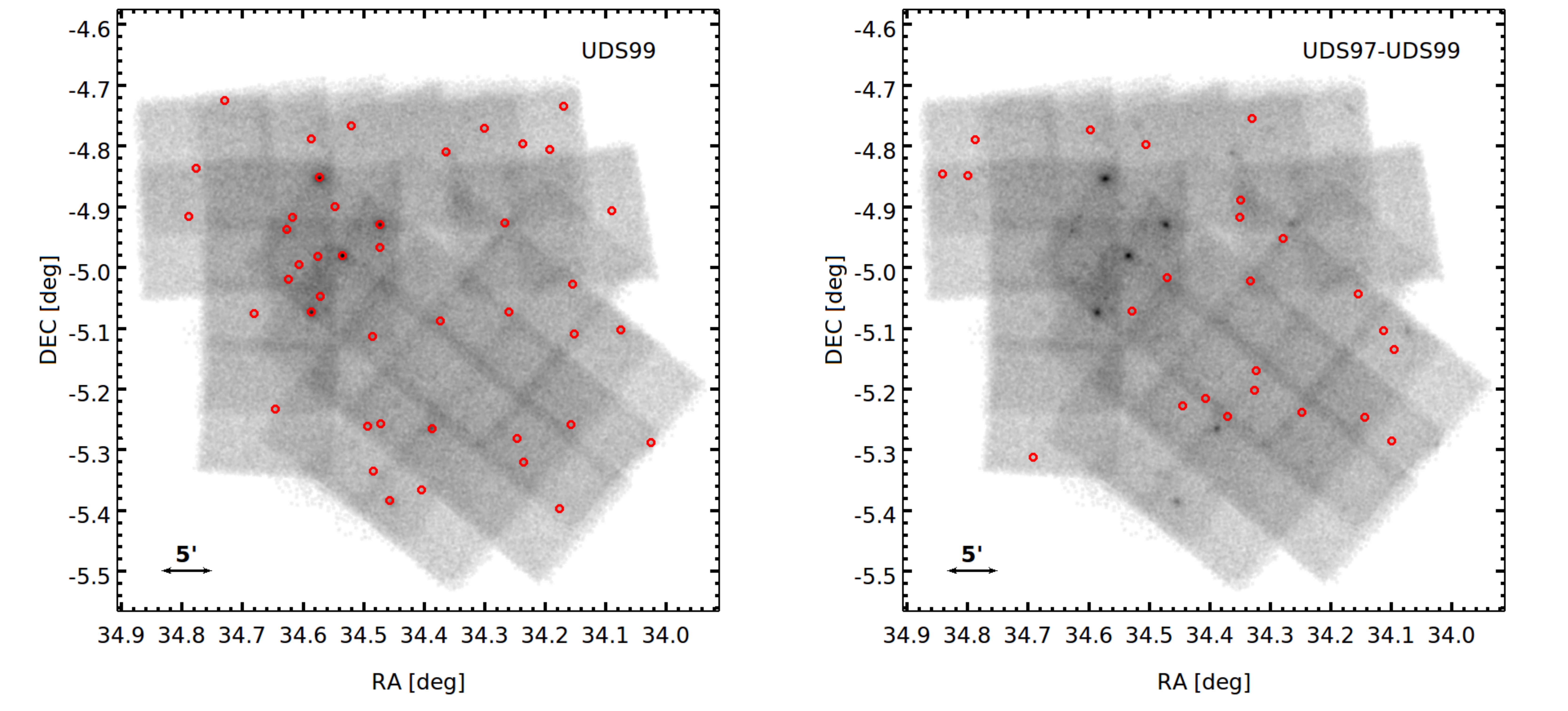}
\caption{\textbf{Left.} NuSTAR $3-24$ keV band mosaic with the 43 sources detected in at least one of the three F, S, H bands (red circles). \textbf{Right.} Same as left panel, but for the 24 additional sources detected in the 97\% reliability catalog, where we detect 67 sources in total. See Section \S \ref{sec:detection} for details. \label{fig:mosaics}}
\end{figure*}

 Following C15, we thaw all the relevant parameters but the normalization of the fCXB, kept frozen to its nominal value \citep{boldt87}. We then fit all the parameters using the Cash statistic \citep{cash79}, and ultimately fit for the fCXB normalization. While this procedure gave very good results in C15 (giving a $< 1\%$ discrepancy in counts between data and background) and in other NuSTAR surveys \citep[e.g.][]{mullaney15}, it underestimates the background counts by $3-4\%$ in the UDS field, which cannot be explained by bright sources, possibly due to a fluctuation of the CXB. After an extensive number of tests, we decide to keep the normalization of the fCXB frozen to its default value \citep{boldt87} in our fits. This recipe, which obviously gives slightly sub-optimal fits in the S band, reconciles our background maps with the data mosaics with a maximum counts discrepancy of $\sim 1\%$.  
Data, exposure, and background mosaics are produced with the FTOOLS task XIMAGE. Data mosaics in the F band, with sources detected above the 99\% and 97\% thresholds of reliability (see the following Sections), are shown in Figure \ref{fig:mosaics}.

\section{Simulations}\label {sec:simul}

In order to define detection parameters, an extensive set (following C15, 400 for each energy band) of simulations is run, following the same strategy adopted by C15, that we briefly summarize here.

\subsection{Initial setup}\label{subsec:setup}

A first run of simulations is performed in each band, distributing sources randomly throughout the FoV and assigning source fluxes from an assumed number counts distribution in the $3-24$ keV band \citep{treister09}, to a minimum flux which is $\sim$ 10 times fainter than the expected survey limit, and placing them on a background without the fCXB component. This is done in order to prevent the simulations from having too many counts, since the fake, unresolved sources make up a part of the fCXB itself. As a result, after the first run of simulations, only a certain fraction of the fCXB contribution is missing. The fraction of fCXB that has to be added to the background depends on the (band-dependent) input limiting flux, and ranges from 61\% in the S band to 94\% in the H2 one. Since between 35 and 55 keV the background is predominantly instrumental, no correction is applied in the VH band. The conversion factors between count rates and fluxes, and the scaling factors from one band to another, adopted throughout this paper, are shown in Table \ref{tab:cf}. \par After rescaling and adding a certain fraction of the fCXB component to the background maps, we run the simulations again and we verify that, on average, our simulations optimally represent our data mosaics. The comparison between observed data and simulated counts in each band is shown in Table \ref{tab:sim}. 

\begin{table}
\caption{Conversion factors from count rates to fluxes used in the paper, from WebPIMMS\tablenotemark{a}, and flux scaling factors with respect to the F band flux, obtained assuming a power law spectrum with $\Gamma = 1.8$.}             
\label{tab:cf}      
\centering          
\begin{tabular}{l c c} 
\hline\hline       
\noalign{\vskip 0.5mm} 
Band & Conversion Factor [erg cm$^{-2}$] & Scale factor \\ 
\noalign{\vskip 0.5mm}   
$3-24$ keV & $4.86 \times 10^{-11}$  & 1.\\ \noalign{\vskip 0.5mm} 
$3-8$ keV & $3.39 \times 10^{-11}$ & 0.42 \\ \noalign{\vskip 0.5mm} 
$8-24$ keV & $7.08 \times 10^{-11}$ & 0.58 \\ \noalign{\vskip 0.5mm} 
$8-16$ keV  & $5.17 \times 10^{-11}$ & 0.35 \\ \noalign{\vskip 1mm} 
$16-24$ keV & $1.62 \times 10^{-10}$ & 0.23 \\ \noalign{\vskip 0.5mm} 
$35-55$ keV &  $1.07 \times 10^{-9}$ & 0.30 \\ \noalign{\vskip 0.5mm}
\noalign{\vskip 1mm}    
\hline
\end{tabular}
\tablenotetext{a}{https://heasarc.gsfc.nasa.gov/cgi-bin/Tools/w3pimms/w3pimms.pl}
\end{table}

\begin{table}
\caption{Comparison between observed counts and average of simulations, after the second run.}             
\label{tab:sim}      
\centering          
\begin{tabular}{l c c c c c c} 
\hline\hline       
\noalign{\vskip 0.5mm} 
 & Data & $\langle \rm Sim\rangle$ & $( \rm D -\langle \rm Sim\rangle)/\rm D$ [\%] \\ 
\noalign{\vskip 0.5mm}  
\cline{2-4}                   
\noalign{\vskip 1mm}  
$3-24$ keV & 873931 & 874376 & $-0.05$ \\ \noalign{\vskip 0.5mm} 
$3-8$ keV & 392794  & 392723 & $+0.02$  \\ \noalign{\vskip 0.5mm} 
$8-24$ keV & 481137 & 481356 & $-0.05$  \\ \noalign{\vskip 0.5mm} 
$8-16$ keV  & 291663 &  291646 & $+0.006$   \\ \noalign{\vskip 1mm} 
$16-24$ keV & 189474  & 189513  & $-0.02$ \\ \noalign{\vskip 0.5mm} 
$35-55$ keV\tablenotemark{a} & 357133  & 358531 & $-0.4$ \\ \noalign{\vskip 0.5mm}
\noalign{\vskip 1mm}    
\hline
\end{tabular}
\tablenotetext{a}{We do not rescale the fCXB component for the VH band, since the background is predominantly instrumental.}
\end{table}

\subsection{Reliability, completeness, and sensitivity}\label{subsec:rel}

Once the simulations are completed, we have a set of 2400 simulations (400 for each band) which on average accurately represent our real observations. This large set of simulations is used to maximize the efficiency of our detection procedure. 
\par Following C15, we smooth every simulation (and the background mosaic) with 10$\arcsec$ and 20$\arcsec$ radii circular top-hat functions in order to separate close sources with the first scale, and to detect faint sources with the second one. Then, we convert the resulting smoothed maps into probability maps, using the \texttt{igamma} function (i.e. incomplete $\Gamma$ function) in IDL, which returns the probability of having a certain number of counts $C_{\rm im}$ in the data mosaic given $C_{\rm bkg}$ background counts in the background mosaic at the same position: 

\begin{equation}
P = \texttt{igamma}(C_{\rm im},C_{\rm bkg}).
\end{equation}

In every point of our probability maps, the numerical value is then given by $\log{1/P}$. We use the \textit{SExtractor} software \citep{bertin96} on the probability maps (both 10$\arcsec$ and 20$\arcsec-$smoothed) to detect sources in our simulations. The two lists of sources are merged together; as discussed in C15, using two different smoothing radii increases the number of detections. Every source is then evaluated calculating its Poisson probability of being a spurious fluctuation of the background. Every source is assigned a DET\_ML number, which is simply DET\_ML $=-\ln{P}$. The higher the DET\_ML, the higher the probability of the source being real, and the higher its significance. In the case where the same source is found in both the 10$\arcsec$ and 20$\arcsec-$smoothed maps, the more significant one (i.e., the one with the highest DET\_ML) is retained. Duplicates are assessed cross-correlating the catalogs with a matching radius of 30$\arcsec$, which is found by C15 to better take into account the tail of faint sources matched to their counterparts \citep[see Figure 4 of][]{civano15}. 
Following \citet{mullaney15}, a deblending algorithm for counts of the detected sources is run, in order to take into account the possible contaminations induced by objects closer than 90$\arcsec$. A deblended DET\_ML is then re-calculated using deblended source and background counts to assess the post-deblending significance of every source.
\par
\begin{figure}
\plotone{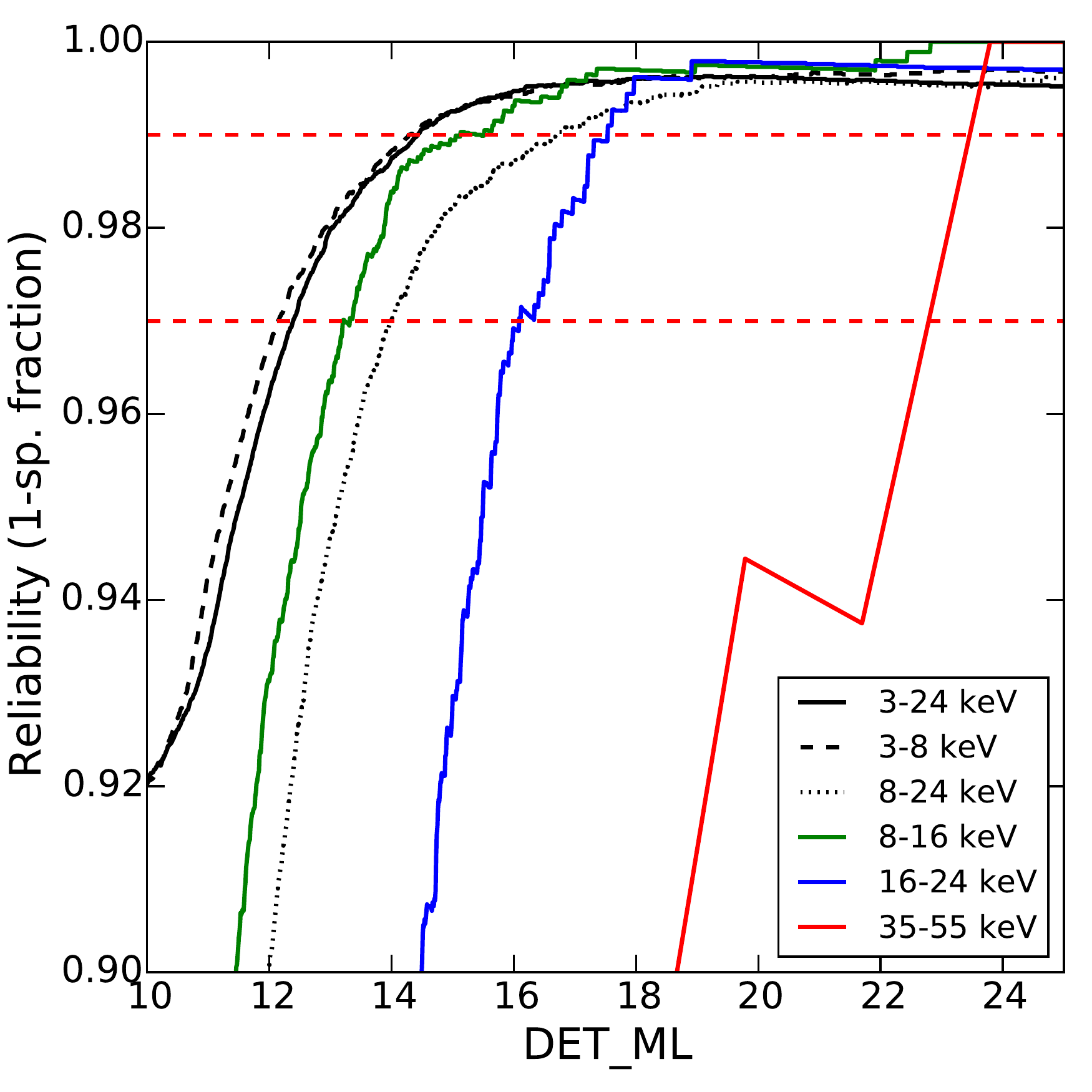}
\plotone{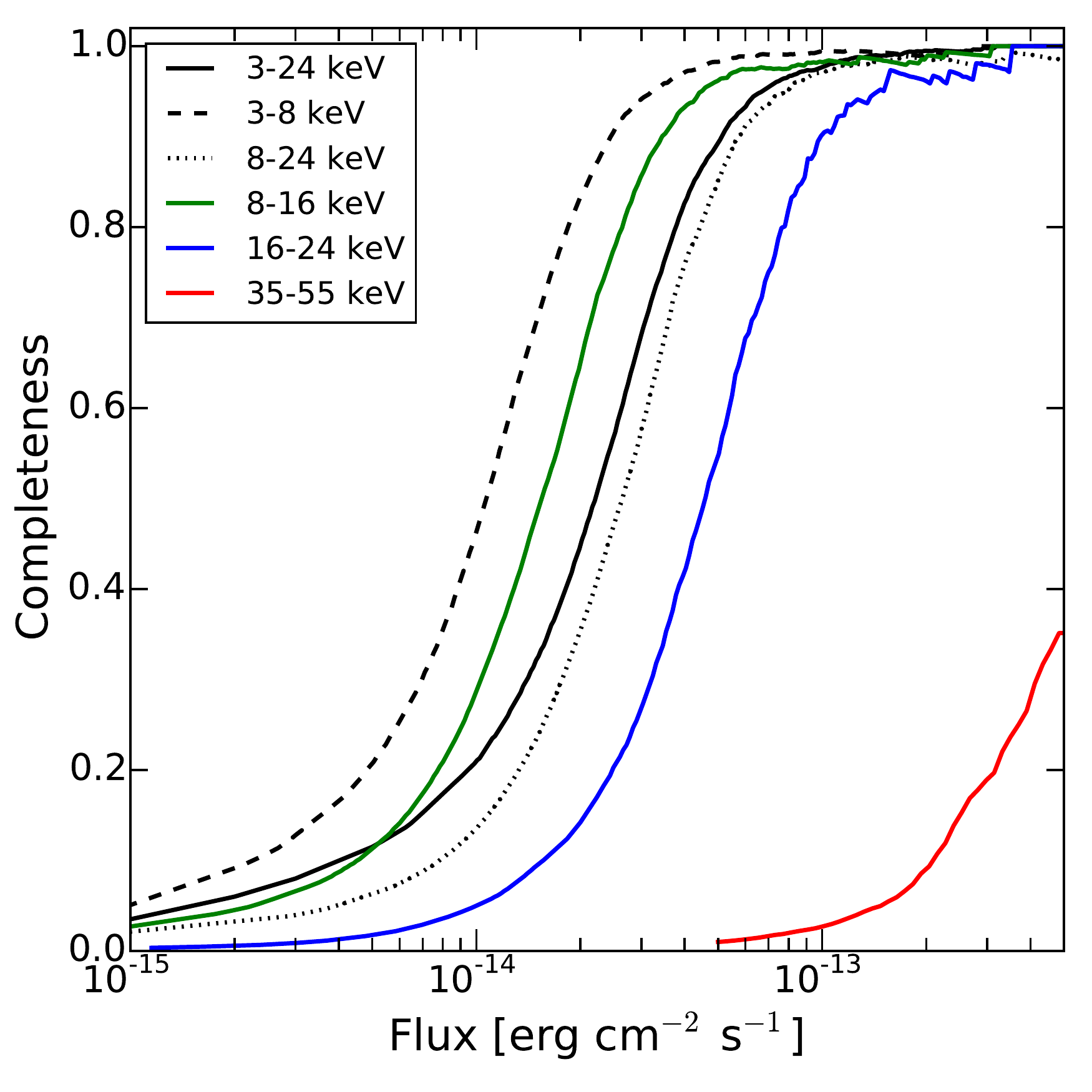}
\plotone{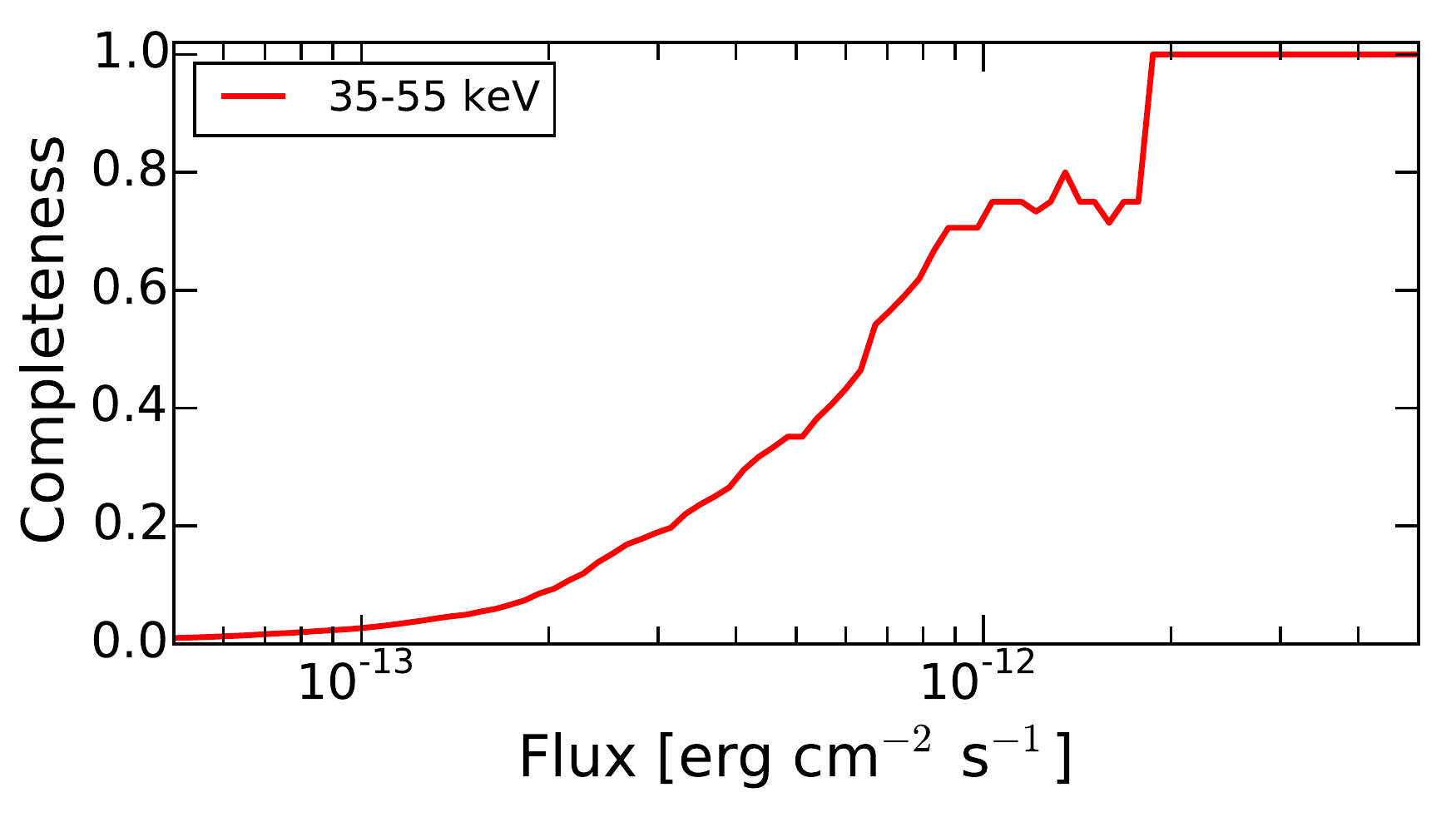}
\caption{\textbf{Top.} Cumulative reliability as a function of DET\_ML. The two red dashed horizontal lines show the 99\% and 97\% of reliability thresholds. In black, the three canonical bands are shown as solid (F), dashed (S), and dotted (H) lines. In green, blue and red solid lines we show the H1, H2 and VH bands, respectively. We note that the harder the band, the noisier the curve, due to low statistics. \textbf{Middle.} Cumulative completeness as a function of input flux, at the 97\% of reliability. Colors and line styles are the same as in the left panel. The completeness curve for the VH band is barely shown because it lies out of scale, which is chosen to emphasize the differences between the canonical bands instead. \textbf{Bottom.} Same of middle panel, but for the VH band.\label{fig:compl}}
\end{figure}

 After these steps, we end up with a catalog of sources for every simulation. Comparing the final list of sources, detected and matched, with the ones input to the simulations allows the calculation of the sample reliability, i.e. the ratio between the cumulative distribution of matched sources and the cumulative distribution of detected sources, as a function of their significance:
\begin{equation}
\text{Rel}(\rm DET\_ML) = \frac{\rm Matched}{\rm Detected}.
\end{equation}
Highly significant sources are also correctly matched to their input counterparts, and the reliability curve is unity at high values of DET\_ML. It then falls steeply at lower significance, where the number of spurious detections starts to increase. We can set a DET\_ML threshold where the reliability falls to the 99\%, or 97\% of its maximum value; at these thresholds, we expect to have a spurious fraction of 1\% and 3\%, respectively. As an example, in the full 3--24 keV band, these thresholds (DET\_ML = 14.42  and DET\_ML = 12.39) correspond to a probability P $\sim 5.5 \times 10^{-7}$ and P $\sim 4.1 \times 10^{-6}$, respectively, of a source being spurious. The top panel of Figure \ref{fig:compl} shows the cumulative distribution of reliability for all our bands.
\par
\begin{figure}
\plotone{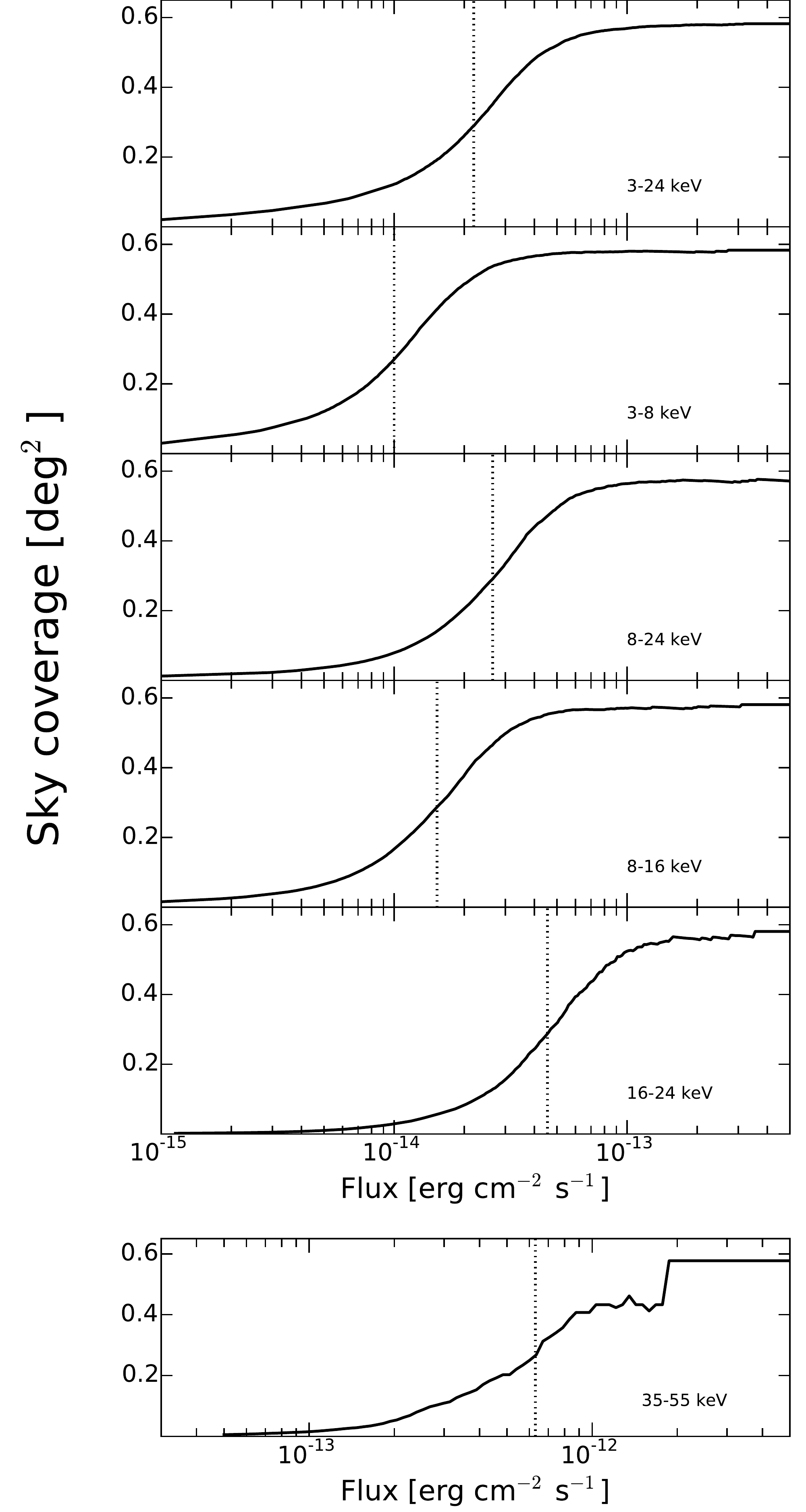}
\caption{Sky coverage as a function of input flux for all the bands (from top to bottom: F, S, H, H1, H2, VH). In each panel, the dotted vertical line marks the half-area flux. Notice the different scale for the VH band in the bottom plot. No area is seen in the VH band at fluxes of $F_{35-55} < 10^{-13}$ erg cm$^{-2}$ s$^{-1}$; very bright sources are needed in order to be detectable in this band. Furthermore, the curve is very noisy due to scarce statistics. Increasing the number of simulations would increase accordingly the number of significantly detected sources and the smoothness of the curve. \label{fig:sens}}
\end{figure}

 Once the DET\_ML threshold is fixed at a given reliability, comparing how many sources are detected above the chosen threshold and matched to the input ones as a function of input flux gives the catalog completeness:
\begin{equation}
\text{Compl}(F_{\rm Input}) = \frac{\text{Detected above thr \& Matched}}{\rm Input}.
\end{equation}
The middle panel of Figure \ref{fig:compl} shows the sample completeness at 97\% reliability for all bands but the VH one, for which the curve is partially shown because it lies off-scale. The completeness curve for the VH band is instead shown in the bottom panel of Figure \ref{fig:compl}. Table \ref{tab:comp} shows different values of completeness for each band. Rescaling the completeness curve for the maximal area of the survey in a given energy band results in a sky coverage, or sensitivity plot. The sensitivities of our survey are shown in Figure \ref{fig:sens}, while Table \ref{tab:simuldetection} summarizes the results of the detection on simulations. Table \ref{tab:simuldetection} shows that the simulations predict non-detection in the VH band, while on average only $1-2$ sources are expected in the H2 band, and the H1 band returns a number of sources larger, or comparable to, the H band. We also note that combining the detection catalogs coming from differently smoothed maps is advantageous only in the hard bands. This is probably due to the fact that the longer scale smoothing improves the sensitivity to faint sources in the hard bands more than in the soft and full ones. \par Finally, we note that, given our assumption of a single photon index of $\Gamma = 1.8$ for the whole AGN population (see \S\ref{subsec:mosaics}), the completeness estimates to recover an \textit{intrinsic} CT fraction may be biased. However, we are going to focus on the \textit{observed} CT fraction throughout this paper.
\par
\begin{figure}
\plotone{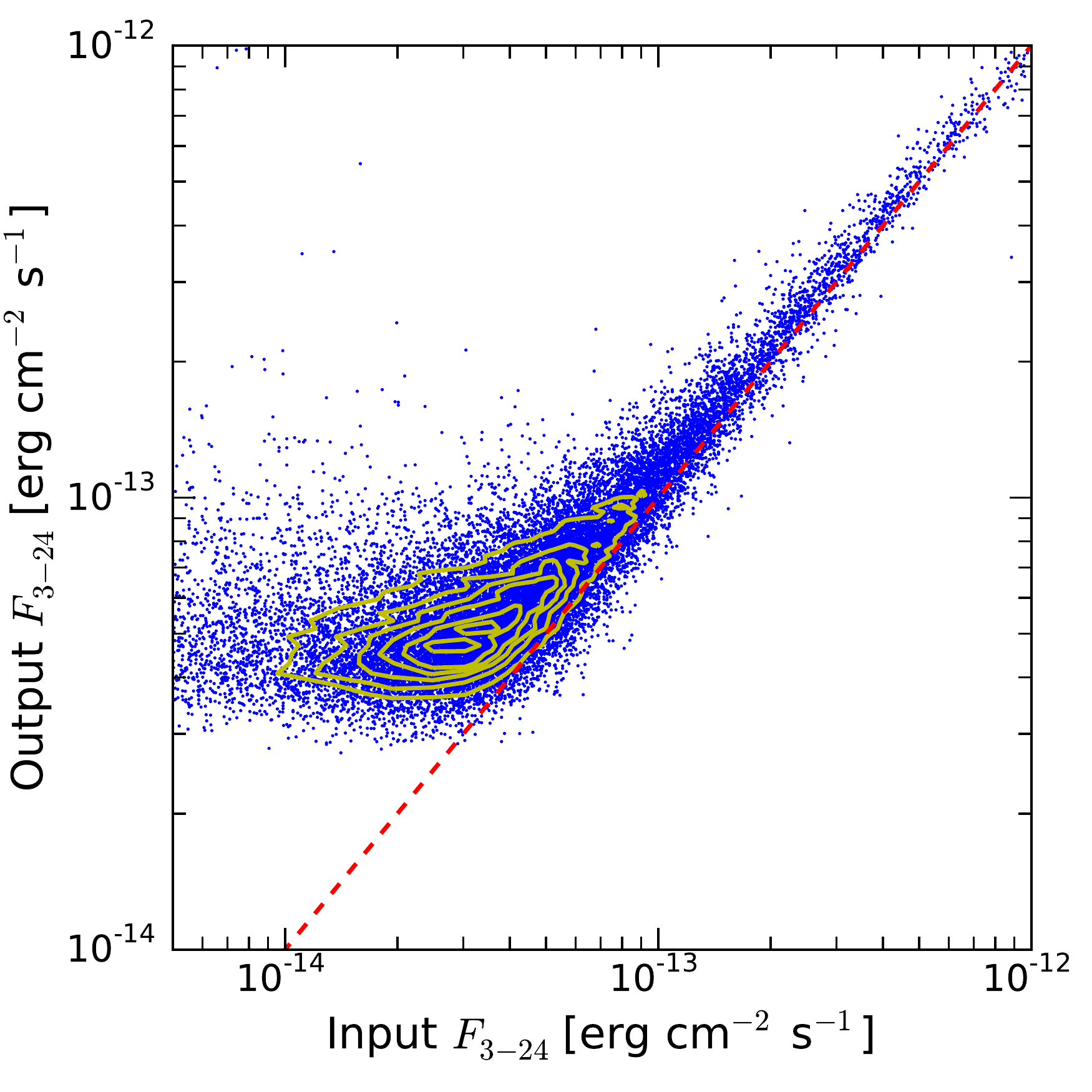}
\plotone{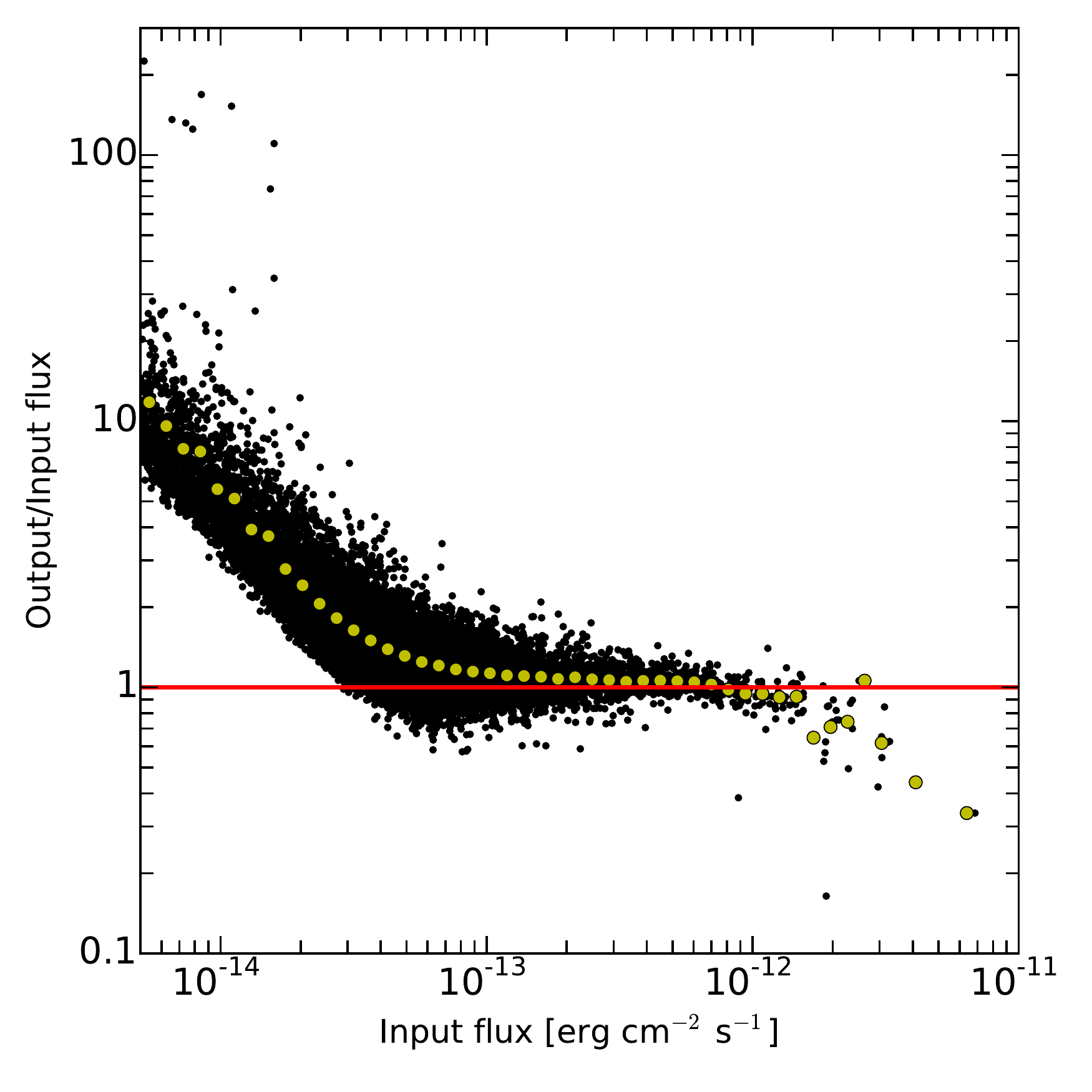}
\caption{\textbf{Top.} Results of aperture photometry on simulations, in the full band. Sources detected above the 97\% reliability threshold and matched to their input counterparts are plotted as blue dots. The red dashed line is the 1:1 relation, while the yellow contours indicate where most of the points lie. The Eddington bias flattening is clearly visible for input fluxes $F_{\rm in} \lesssim 5\times10^{-14}$ erg cm$^{-2}$ s$^{-1}$. \textbf{Bottom.} Ratio between the output and input flux as a function of the input flux for the simulations performed. The yellow dots are averages computed in bins of input flux, while the red horizontal line marks the 1:1 relation. \label{fig:fxfx}}
\end{figure}

 We perform aperture photometry, on the un-smoothed simulations, of detected and matched sources, extracting counts in 20$\arcsec$ circular apertures with standard tools, such as the CIAO task \texttt{dmextract}, and converting the count rates to fluxes with an appropriate, band-dependent, conversion factor (see Table \ref{tab:cf}). We further apply an aperture correction, to get the total flux in each band, such that $F_{\rm corr} = F/0.32$ (see C15 for further details). A direct comparison with the fluxes input to the simulations for the F band is shown in the top panel of Figure \ref{fig:fxfx}, where a good agreement between output and input fluxes is recovered to $F_{\rm in} \sim 5\times 10^{-14}$ erg cm$^{-2}$ s$^{-1}$, where the Eddington bias \citep{eddington13,wang04} makes the relation flatten below $F_{\rm in} \sim 3-4\times 10^{-14}$ erg cm$^{-2}$ s$^{-1}$, corresponding to the flux limit of the survey at $\sim 80\%$ of completeness (see Table \ref{tab:comp}). A small deviation from the 1:1 relation is present also at higher fluxes. This is due to the fixed aperture used to extract NuSTAR counts, from which underestimated fluxes are then computed. In particular, from the simulations we calculate that fluxes above $F_{\rm in} \gtrsim 1.5 \times 10^{-12}$ erg cm$^{-2}$ s$^{-1}$ will be underestimated by $\gtrsim 10\%$. Only $\sim0.1\%$ of the simulated sources fall above this flux threshold, and therefore we do not correct for this bias in our photometry. On the other hand, at an input flux of $F_{\rm in} \sim 10^{-13}$ erg cm$^{-2}$ s$^{-1}$ the output flux is on average overestimated by $\sim13\%$ due to the rising Eddington bias. This is evident in the bottom panel of Figure \ref{fig:fxfx}, where the ratio of output/input flux is shown as a function of the input flux. The yellow dots are the binned averages of the distribution, and help to guide the eye.

\begin{table*}
\caption{Completeness as a function of flux, 97\% reliability catalog. Fluxes are in units of erg cm$^{-2}$ s$^{-1}$.}             
\label{tab:comp}      
\centering          
\begin{tabular}{l c c c c c c} 
\hline\hline       
\noalign{\vskip 0.5mm} 
Completeness & F($3-24$ keV)& F($3-8$ keV) & F($8-24$ keV) & F($8-16$ keV) & F($16-24$ keV)& F($35-55$ keV) \\ 
\noalign{\vskip 0.5mm}   
90\% & $5.1 \times 10^{-14}$  & $2.5 \times 10^{-14}$ & $5.7 \times 10^{-14}$ & $3.5 \times 10^{-14}$ & $9.8 \times 10^{-14}$ & $1.8 \times 10^{-12}$ \\ \noalign{\vskip 0.5mm} 
80\% & $3.8 \times 10^{-14}$ & $1.9 \times 10^{-14}$ & $4.4 \times 10^{-14}$ & $2.6 \times 10^{-14}$ & $7.7 \times 10^{-14}$ & $1.3 \times 10^{-12}$ \\ \noalign{\vskip 0.5mm}
50\% & $2.2 \times 10^{-14}$ & $1.0 \times 10^{-14}$ & $2.7 \times 10^{-14}$ & $1.5 \times 10^{-14}$ & $4.6 \times 10^{-14}$ & $6.3 \times 10^{-13}$ \\ \noalign{\vskip 0.5mm} 
20\%  & $9.5 \times 10^{-15}$ & $5.0 \times 10^{-15}$ & $1.4 \times 10^{-14}$ & $7.5 \times 10^{-15}$ & $2.5 \times 10^{-14}$ & $3.1 \times 10^{-13}$ \\ \noalign{\vskip 0.5mm} 
\noalign{\vskip 1mm}    
\hline
\end{tabular}
\end{table*}

\begin{table*}
\caption{Summary of detections on simulations and real data.}             
\label{tab:simuldetection}      
\centering          
\begin{tabular}{l c c c c c c c c c c} 
\hline\hline       
\noalign{\vskip 0.5mm}  
Simulations & \multicolumn{6}{c}{Bands} \\ \noalign{\vskip 0.5mm} 
 & $3-24$ keV & $3-8$ keV & $8-24$ keV & $8-16$ keV &  $16-24$ keV & $35-55$ keV \\ 
\noalign{\vskip 0.5mm}  
\cline{2-7}                   
\noalign{\vskip 1mm}
10'' smoothed maps & 115 & 97 & 68 & 64 & 33 & 29\\ \noalign{\vskip 0.5mm} 
20'' smoothed maps & 100  & 84 & 57 & 54 & 22 & 15\\ \noalign{\vskip 0.5mm} 
Combined, no duplicates & 115 & 96 & 76 & 70 & 43 & 37\\ \noalign{\vskip 0.5mm} 
Matched to input  & 90 (78\%) &  76 (79\%) & 48 (63\%) & 45 (64\%) & 14 (33\%) & 7 (19\%)\\ \noalign{\vskip 1mm} 
$\rm{DET\_ML}(99\%)$ Thr. & 14.42  & 14.28  & 16.69 & 15.13 & 17.54 & 23.55 \\ \noalign{\vskip 0.5mm}
$\rm{DET\_ML}(97\%)$ Thr. & 12.39  & 12.15  & 14.00 & 13.23 & 16.09 & 23.00 \\ \noalign{\vskip 0.5mm}  
$\rm{DET\_ML} > \rm{DET\_ML}(99\%)$ & 42  & 34  & 13 & 15 & 1 & 0\\ \noalign{\vskip 0.5mm} 
$\rm{DET\_ML} > \rm{DET\_ML}(97\%)$ & 55  & 45 & 19  & 19 & 2 & 0\\ \noalign{\vskip 0.5mm}
\noalign{\vskip 1mm}    
\hline
\noalign{\vskip 1mm}
Real data & & & & & &  \\
$\rm{DET\_ML} > \rm{DET\_ML}(99\%)$ & 40  & 28  & 15 & 16 & 1 & 0\\ \noalign{\vskip 0.5mm} 
$\rm{DET\_ML} > \rm{DET\_ML}(97\%)$ & 61 & 44 & 19  & 21 & 1 & 0\\ \noalign{\vskip 0.5mm}
\noalign{\vskip 1mm}    
\hline
\end{tabular}
\tablecomments{The first (second) row is the average number of sources detected in the 10$\arcsec$ (20$\arcsec$) smoothed maps. The third row is the average number of sources in the merged catalog, cleaned from duplicates. The fourth row is the average number of sources detected and matched (after deblending) to input sources. The fifth and sixth rows display the DET\_ML thresholds at the 99\% and 97\% of reliability. The seventh and eighth rows show the average number of sources expected to be above the 99\% and 97\% reliability thresholds, from the simulations. The last two rows report the effective number of sources detected above the 99\% and 97\% reliability thresholds, respectively.}
\end{table*}

\section{Source detection} \label{sec:detection}

\begin{figure}
\plotone{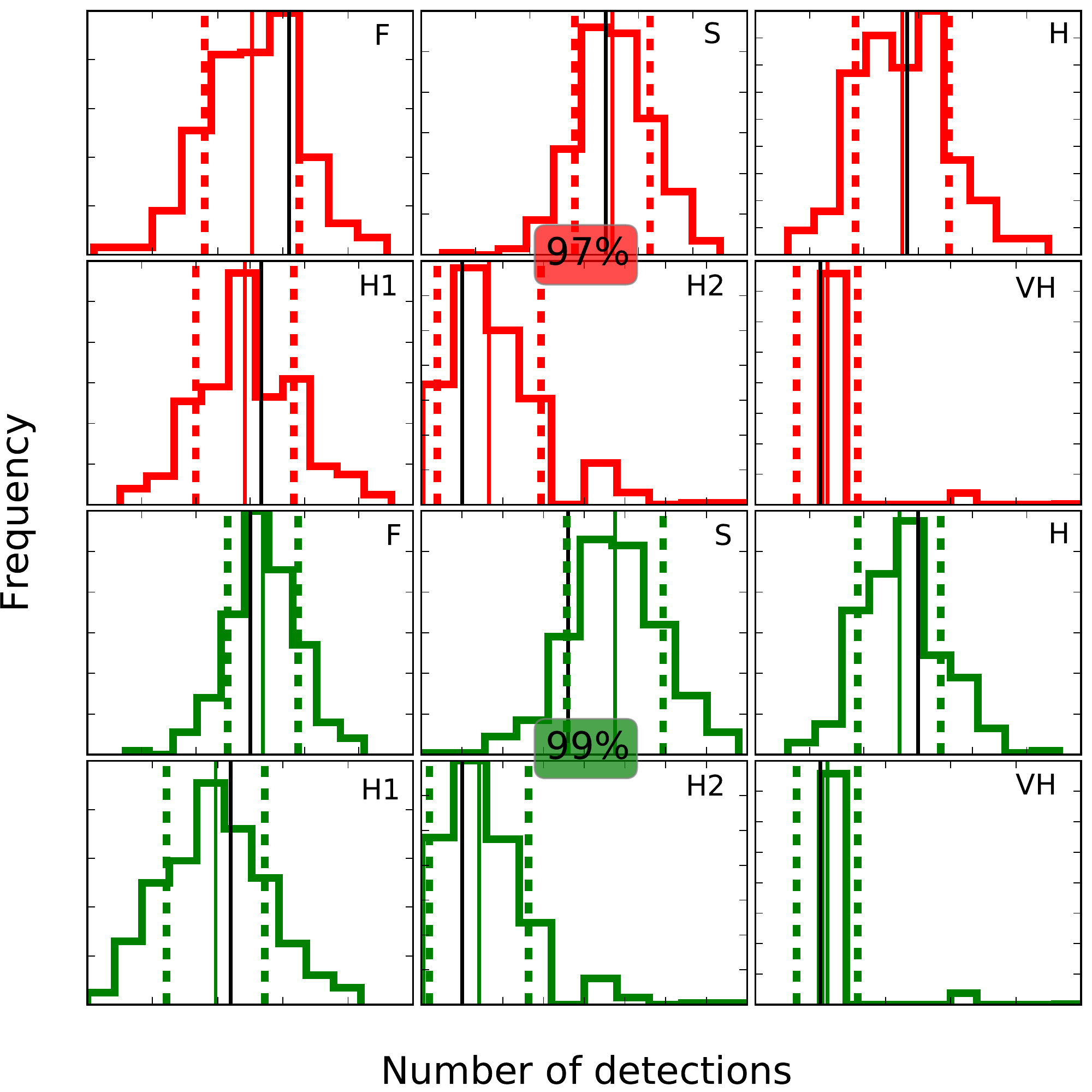}
\caption{Distribution of detections within the simulations. The six upper panels, in red, refer to the six bands at 97\% of reliability threshold. The six lower panels, in green, to the 99\% of reliability. From top left to bottom right, the bands are: F, S, H, H1, H2, and VH. In each panel, the dashed lines mark the $\pm 1\sigma$ interval from the average, while the black solid line marks the position of the detections in the data mosaic, which are always within $\pm1\sigma$ from the mean of the simulations, shown by the solid red or green line. Numerical values for the average number of detections in the simulations in each band, and the number of detections in the data mosaics, are explicitly addressed in the last four rows of Table \ref{tab:simuldetection}. \label{fig:distrib}}
\end{figure}

We repeat the same procedure of source detection on the data mosaics. After deblending the list of potential sources, we detect 43 unique sources above the threshold of 99\% of reliability in at least one of the ``canonical'' F, S, and H bands. When considering the 97\% reliability threshold, we detect 67 sources. We will refer to these catalogs as UDS99 and UDS97, respectively, and the detailed numbers for each band are reported in the last two rows of Table \ref{tab:simuldetection}. We note that the numbers of detections agree very well (within 1$\sigma$ of the distributions; see Figure \ref{fig:distrib}) with the simulation expectations. To maximize the statistics, we will focus on UDS97 henceforth, keeping in mind that the spurious fraction of this catalog is 3\%\footnote{Details about this catalog can be found in Appendix \ref{sec:catalog}.}. For homogeneity with the other NuSTAR surveys, and given the few detections in the H1, H2 and VH bands, we will also consider only the 67 sources detected in the canonical bands (of which two are expected to be spurious), while discussing the results in the new bands elsewhere (Masini et al., in prep). Table \ref{tab:detection} shows how these 67 sources are distributed within the F, S and H bands.

\begin{table}
\caption{Summary of detection, UDS97.}             
\label{tab:detection}      
\centering          
\begin{tabular}{l r} 
\hline\hline       
\noalign{\vskip 0.5mm}  
 Bands & Number of sources \\ 
\noalign{\vskip 0.5mm} 
\hline                   
\noalign{\vskip 1mm}  
F + S + H &	14 (21\%)\\ \noalign{\vskip 0.5mm}
F + S + h &	10 (15\%) \\ \noalign{\vskip 0.5mm}
F + s + H &	1 (2\%)\\ \noalign{\vskip 0.5mm}
F + S	 &	15 (22\%)\\ \noalign{\vskip 0.5mm}
F + s &		3 (4\%) \\ \noalign{\vskip 0.5mm}
f + S &		4 (6\%)\\ \noalign{\vskip 0.5mm}
F + H &		3 (4\%)\\ \noalign{\vskip 0.5mm}
F + h &		8 (12\%)\\ \noalign{\vskip 0.5mm}
f + H &		1 (2\%)\\ \noalign{\vskip 0.5mm}
F &			7 (10\%)\\ \noalign{\vskip 0.5mm}
S &			1 (2\%)\\ \noalign{\vskip 0.5mm}
\noalign{\vskip 1mm}    
\hline              
\end{tabular}
\tablecomments{Capital letters are for sources above the threshold, while lower case letters refer to sources detected but below the threshold.}
\end{table}

\begin{figure*}
\plottwo{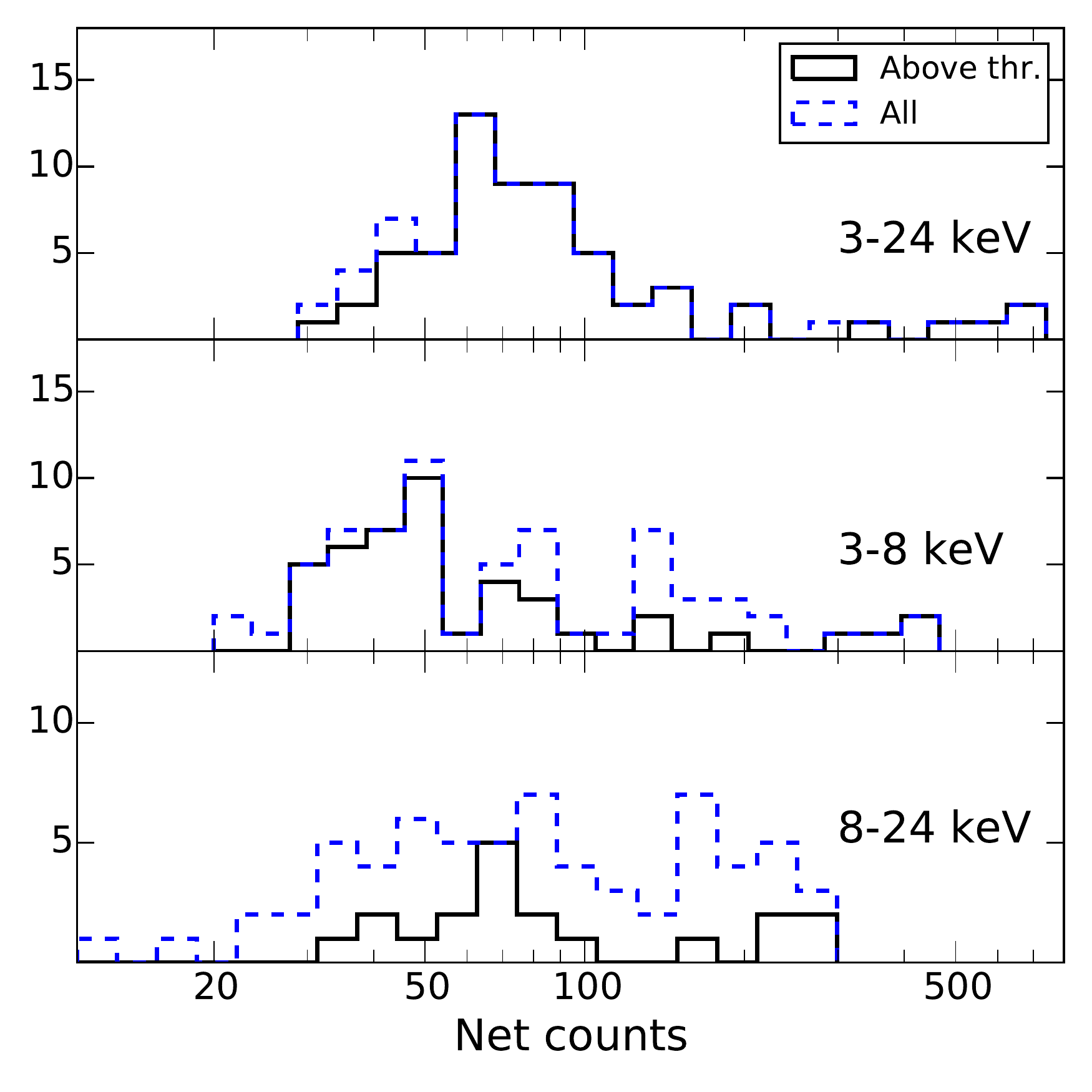}{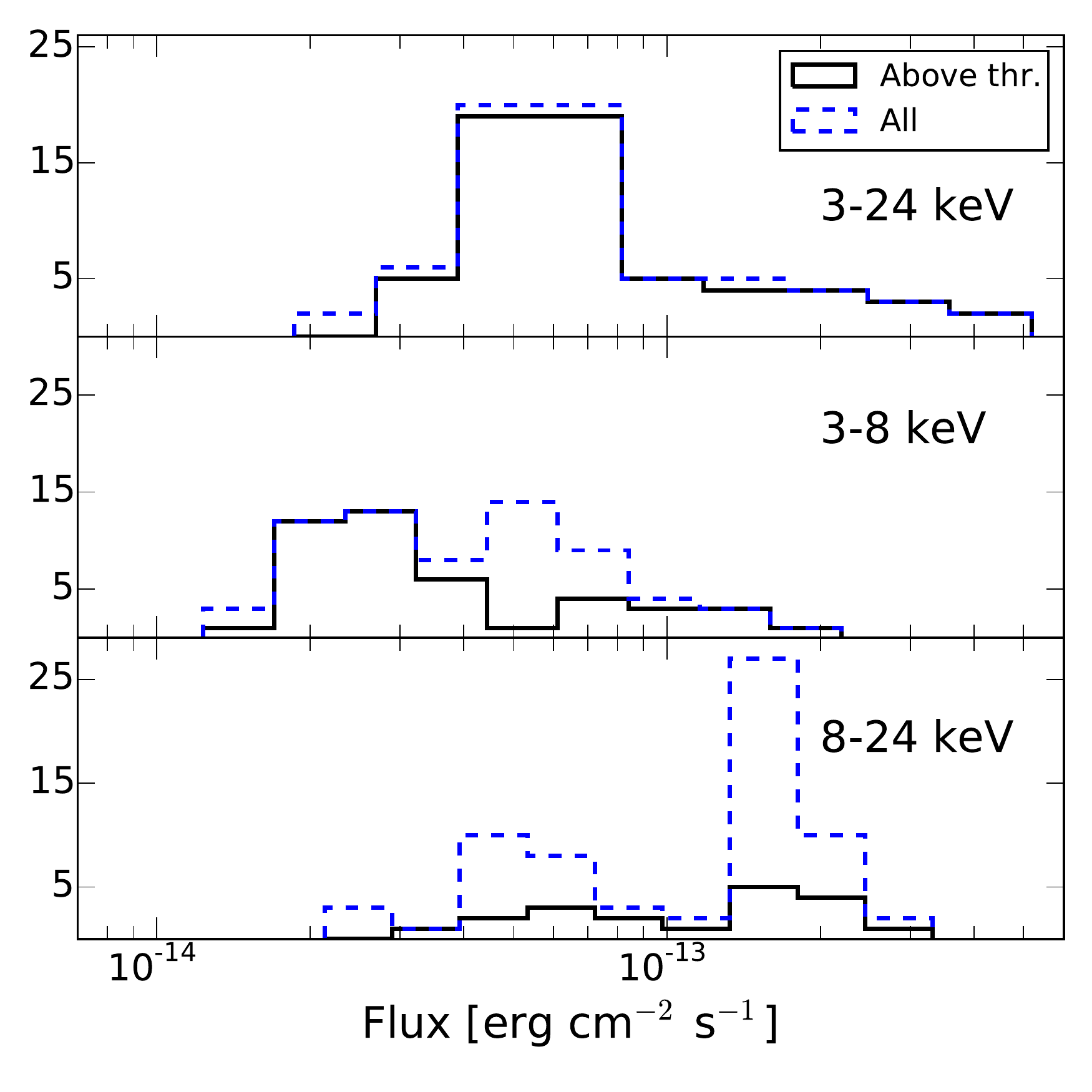}
\caption{\textbf{Left.} Counts distribution in the F band (top panel), S band (middle panel) and H band (bottom panel) for all the sources (blue dashed histogram), and sources above the threshold only (black solid histogram). Note that the ``All'' histograms (dashed blue lines) include upper limits. \textbf{Right.} Same as left panel, but for the fluxes.\label{fig:counts}}
\end{figure*}

\subsection{Catalog creation}\label{subsec:catalog}

We create a catalog of the 67 sources detected in the three canonical bands of UDS97. Following the same strategy adopted for the simulations, we use \texttt{dmextract} to get total and background counts from circles of 20$\arcsec-$radius, from the data and background mosaics, respectively. Similarly, we extract the exposures from the exposure map mosaic in order to compute the count rates for each source and each band, which are then aperture-corrected and converted into fluxes with the appropriate conversion factors. If a source is detected below the threshold, or undetected, in a given band, $3\sigma$ upper limits are provided extracting counts at its position. For detections, 1$\sigma$ uncertainties are obtained using Equations 9 and 12 of \citet{gehrels86} with $S = 1$, while for non-detections we use Equation 10 of \citet{gehrels86} with $S = 3$. The distributions of net counts and fluxes for our sources are shown in Figure \ref{fig:counts}.

\section{Match with \xmm\ and \chandra\ catalogs}\label{sec:match}

The 67 NuSTAR-detected sources are cross-matched with the Subaru \xmm\ Deep Survey catalog of  \cite{ueda08} and \chandra\ catalog of Kocevski et al. (submitted), with a matching radius of 30$\arcsec$. A flux cut was applied to both catalogs, excluding counterparts with a $3-8$ keV flux more than a factor of three lower than the NuSTAR flux limit at the 50\% of completeness in the $3-8$ keV band (i.e., $F_{\rm cut} \sim 3 \times 10^{-15}$ erg cm$^{-2}$ s$^{-1}$). 	
Any other soft X-ray counterpart below this flux cut is at least at $2.7\sigma$ and $1.9\sigma$ from the NuSTAR flux for \xmm\ and \chandra, respectively.

\subsection{\xmm}\label{sec:match_xmm}

We directly match 88\% of our sources (59/67) with \xmm\ sources in the SXDS catalog. We further find a counterpart at a distance of 4$\arcsec$ for uds7, which falls in the tiny fraction of NuSTAR area not covered by the SXDS survey (see Figure \ref{fig:uds_grid}), in the 3XMM$-$DR6 catalog \citep{rosen16} so that the final fraction of \xmm-matched sources is 60 out of 67 (90\%). Of the seven sources not matched, uds59 is detected above the 97\% threshold in the F, S and  H bands. It is also above the threshold in the H1 band, and is detected above the 99\% reliability threshold in the F, H, and H1 bands. More details on this source are provided in \S \ref{sec:uds59}. Two other sources have an \xmm\ counterpart, albeit undetected by \xmm\ in the $4.5-10$ keV band from which the $3-8$ keV flux is computed, while one is a blending of two SXDS sources with the same $3-8$ keV flux, thus slightly lower than the chosen threshold, and are then excluded by our cut. Even if included, this last source would have been considered as a blending, and therefore excluded from the following analysis. The remaining three unassociated sources could be the spurious ones expected from our chosen reliability threshold.
Out of 60 matched sources, four have two possible \xmm\ counterparts, and one has triple counterparts within 30$\arcsec$. We have then 55 sources with unique \xmm\ counterparts. To properly deal with multiple counterparts, we adopt the following strategy: if the closest of the counterparts within 30$\arcsec$ also has the highest ``hard'' (i.e., $3-8$ keV) flux, it is considered as the primary counterpart of the NuSTAR source. Otherwise, it is considered blended and is then excluded from the following analysis. With this prescription, we add two sources (uds2 and uds18) in which the closest \xmm\ counterpart is also the brightest, providing a total of 57 matches. The flux ratios between the primary and secondary counterpart in these two cases are 1.2 for uds2 and 4.3 for uds18.

\subsubsection{The case of uds59}\label{sec:uds59}
As previously mentioned (\S \ref{sec:match}), uds59 is detected by NuSTAR above the 97\% reliability thresholds in four out of six bands (F, S, H, and H1). We note that it is present also in the UDS99 catalog, although it is below the threshold in the S band. Detected with 82 net counts in the F band, uds59 is located in a part of the mosaic where only \xmm\ coverage is available, but it is not matched to any SXDS source. At its position, SDSS maps show the presence of a group of galaxies, which is indeed detected as a galaxy cluster in the CFHTLS 4 Wide Fields Galaxy Clusters catalog \citep{durret11} at $z_{\rm phot} \sim 0.45$. Moreover, two galaxies of the group are detected by WISE \citep{wright10} and are 4.7$\arcsec$ and 11.5$\arcsec$ away from the NuSTAR position. \par Since uds59 is then a strong candidate to be a newly discovered source, we extract its NuSTAR X-ray spectrum, assuming the redshift of the group ($z_{\rm phot} = 0.45$) and we fit the spectrum with a simple Galactic-absorbed power law. The returned photon index is quite flat, implying that the source is obscured ($\Gamma = 0.68^{+0.53}_{-0.54}$), consistent with the non-detection by \xmm. Adding a screen along the line of sight (through a \texttt{zwabs} model) and fixing $\Gamma = 1.8$, we get a good fit (CSTAT/DOF=200/235) with the source being heavily obscured ($N_{\rm H} = 7.4^{+5.8}_{-4.1} \times 10^{23}$ cm$^{-2}$). A very similar result is obtained using a MYTorus model \citep[CSTAT/DOF=202/235, $N_{\rm H} = 6.5^{+5.3}_{-3.7} \times 10^{23}$ cm$^{-2}$]{murphy09}.
We can also use these models to calculate the flux in the three canonical bands, and we have:

\begin{eqnarray}
\nonumber F_{3-24}= 1.2^{+0.3}_{-0.3} \times 10^{-13} \rm ~erg~ cm^{-2} s^{-1}, \\
\nonumber F_{3-8}= 2.3^{+1.0}_{-0.8} \times 10^{-14} \rm ~erg~ cm^{-2} s^{-1}, \\
\nonumber F_{8-24}= 9.7^{+2.9}_{-4.7} \times 10^{-14} \rm ~erg~ cm^{-2} s^{-1}. 
\end{eqnarray}

We can compare the soft band flux with the upper limit obtained from the $0.2-12$ keV \xmm\ mosaic. Extracting the total number of counts in a circular region of 20$\arcsec-$radius and using the average vignetting-corrected exposure at the same position, we get a count rate of $1.16\times10^{-2}$ cts s$^{-1}$ for the PN, which translates into a predicted $F_{3-8} \sim 9.2 \times 10^{-15}$ erg cm$^{-2}$ s$^{-1}$.
A combination of intrinsic obscuration and low-exposure coverage (the effective exposure time at the position of the source is only 9.8 ks for the PN) is likely responsible for the non-detection by \xmm.

\subsection{\chandra}

As $\sim 30\%$ of the NuSTAR UDS field is not covered by \chandra, only 41 sources have \chandra\ coverage, and 40 of them are matched. Notably, uds45, the only one missing a \chandra\ counterpart, is not matched in the \xmm\ catalog either, and is then a strong candidate to be one of the two expected spurious sources in the catalog.
Out of 40 \chandra-matched sources, five have a double counterpart and two have a triplet in the \chandra\ catalog. As done for \xmm, each case is evaluated and we add another four counterparts to the 33 unique matches, for a total of 37 matches. The flux ratios between the primary and secondary/tertiary counterpart range between $\sim 1.5-3.4$.
\par
\begin{figure*}
\plottwo{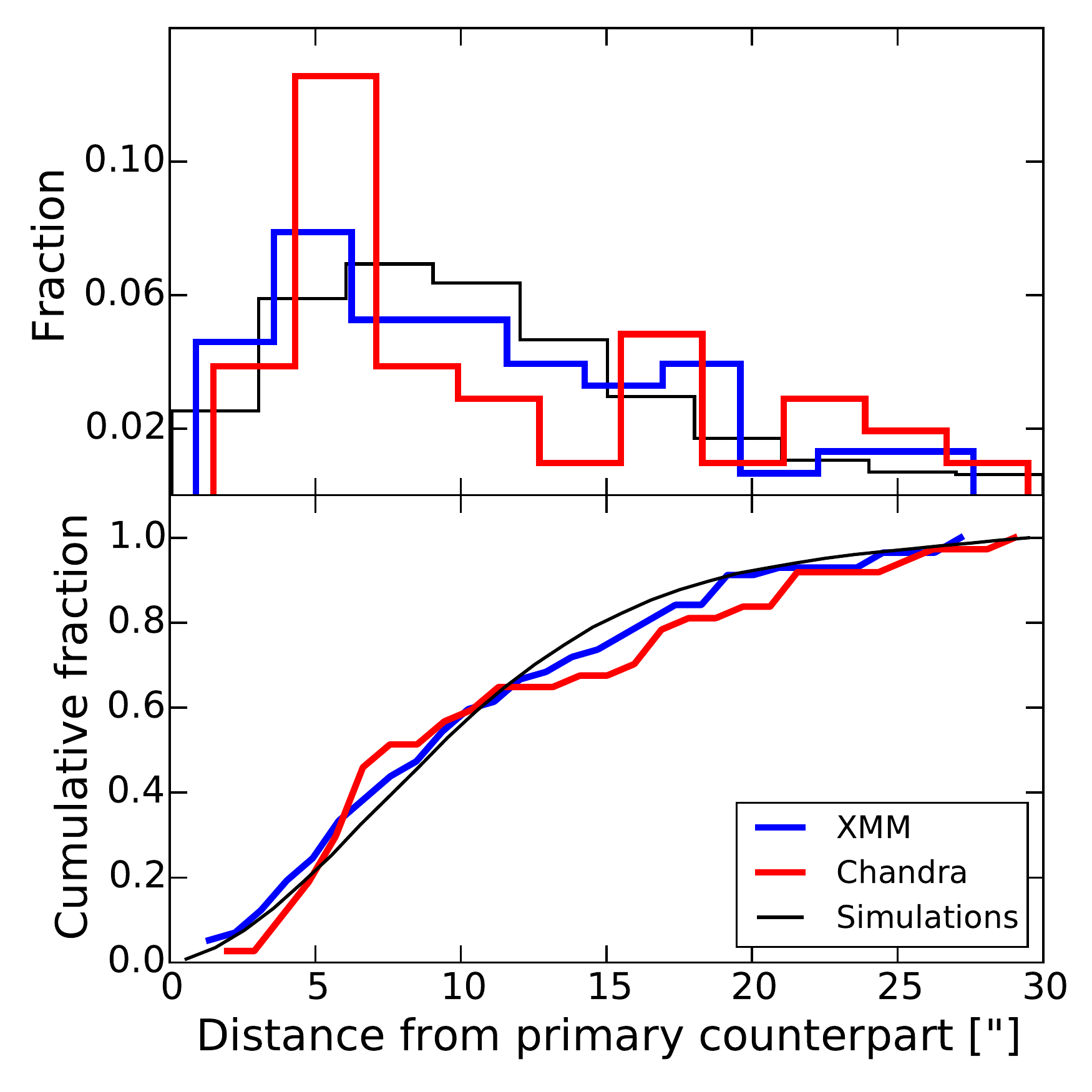}{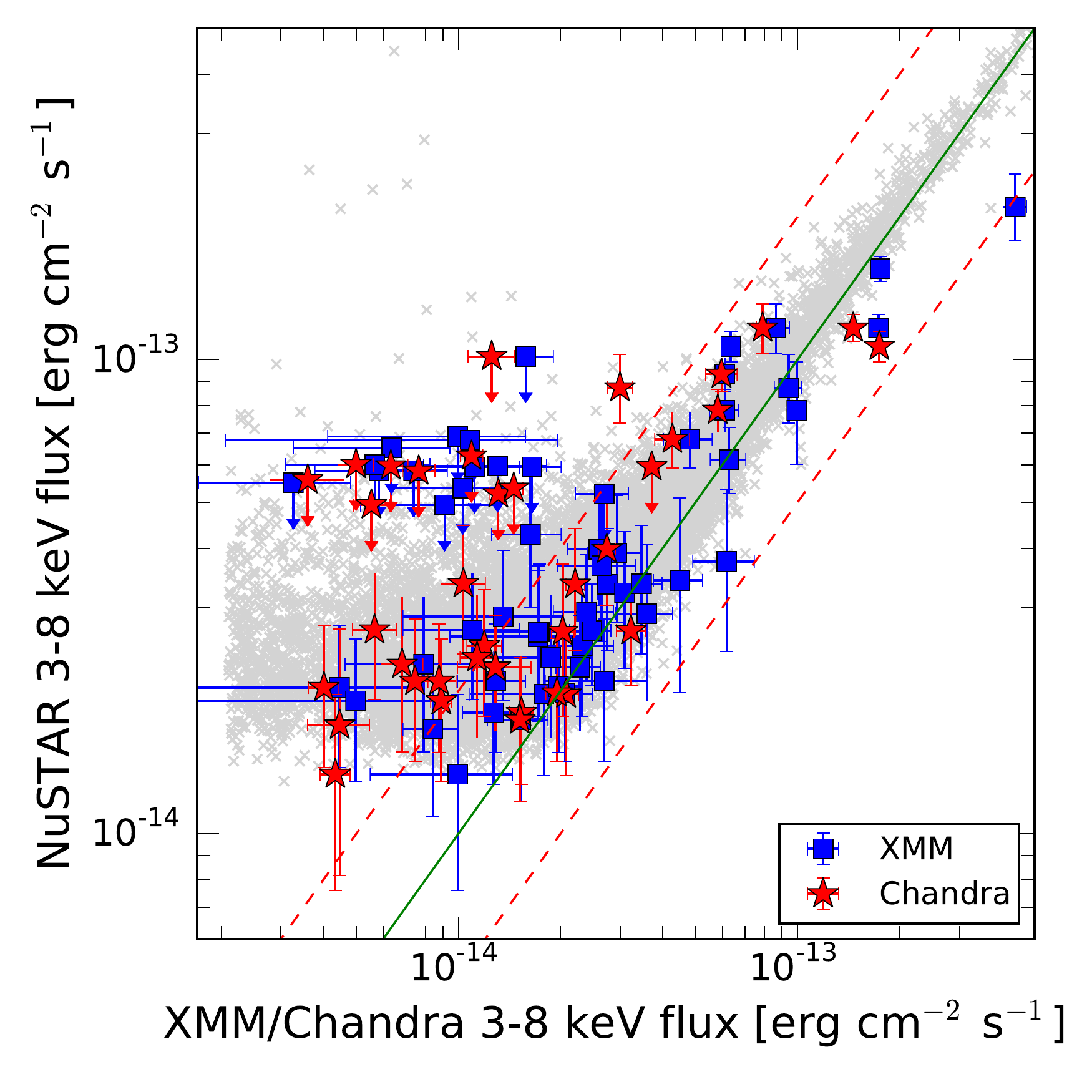}
\caption{\textbf{Left.} (Top) Distribution of separation between our NuSTAR sources  and \xmm\ (blue solid line) and \chandra\ (red solid line) counterparts, compared with the simulations in the S band. (Bottom) Cumulatives of the distributions shown in the top panel. More than 50\% of our sources are matched within 10$\arcsec$, and 80\% within 20$\arcsec$. Colors as in the top panel. \textbf{Right.} NuSTAR $3-8$ keV fluxes as a function of \xmm\ (blue squares) and \chandra\ (red stars) $3-8$ keV fluxes. Upper limits at the 3$\sigma$ confidence level on NuSTAR fluxes are represented as downward arrows. The green solid line is the 1:1 relation, while the red dashed lines are a factor of two displaced from it. At low fluxes, the Eddington bias makes the points deviate from the 1:1 relation, while the displaced point close to the right corner of the plot (uds7) may have an underestimated NuSTAR flux due to its position, on the very edge of the mosaic. The gray crosses on the background are the expectations from the simulations in the $3-8$ keV band (see Figure \ref{fig:fxfx}). \label{fig:fluxes}}
\end{figure*}

 The distribution of separations between our NuSTAR sources and their low-energy counterparts, and its cumulative, are shown in Figure \ref{fig:fluxes} (left panel); 60\% of the sources are matched within 10$\arcsec$, while $80-90\%$ are matched within 20$\arcsec$. We note that the distribution peaks are consistent with the simulated source distribution in the $3-8$ keV band. These fractions are comparable with the ones found by C15 taking into account the secondary counterparts, while slightly lower than the separations with the primary counterparts, but considering a higher reliability sample. In the right panel of the same Figure \ref{fig:fluxes} we also compare NuSTAR, \xmm\ and \chandra\ fluxes in the $3-8$ keV band converting the \xmm\ $4.5-10$ keV count rates and the $2-10$ keV \chandra\ fluxes to $3-8$ keV fluxes assuming a $\Gamma=1.8$ power law. As can be seen from the right panel of Figure \ref{fig:fluxes}, there is some scatter between the fluxes measured by NuSTAR and soft X-ray instruments fluxes, although the NuSTAR fluxes have large uncertainties ($1\sigma$ uncertainties for detections, $3\sigma$ upper limits for non-detections). This scatter, which is increased by the Eddigton bias at the lowest fluxes, while being always less than a factor of two at the brightest fluxes (i.e., at $F \gtrsim 5\times10^{-14}$ erg cm$^{-2}$ s$^{-1}$), can be explained in part by X-ray variability, which is a common property of AGN \citep[e.g.,][]{paolillo17}, and in part by cross-calibration uncertainties.

\subsection{Optical counterparts}
\begin{figure*}
\plottwo{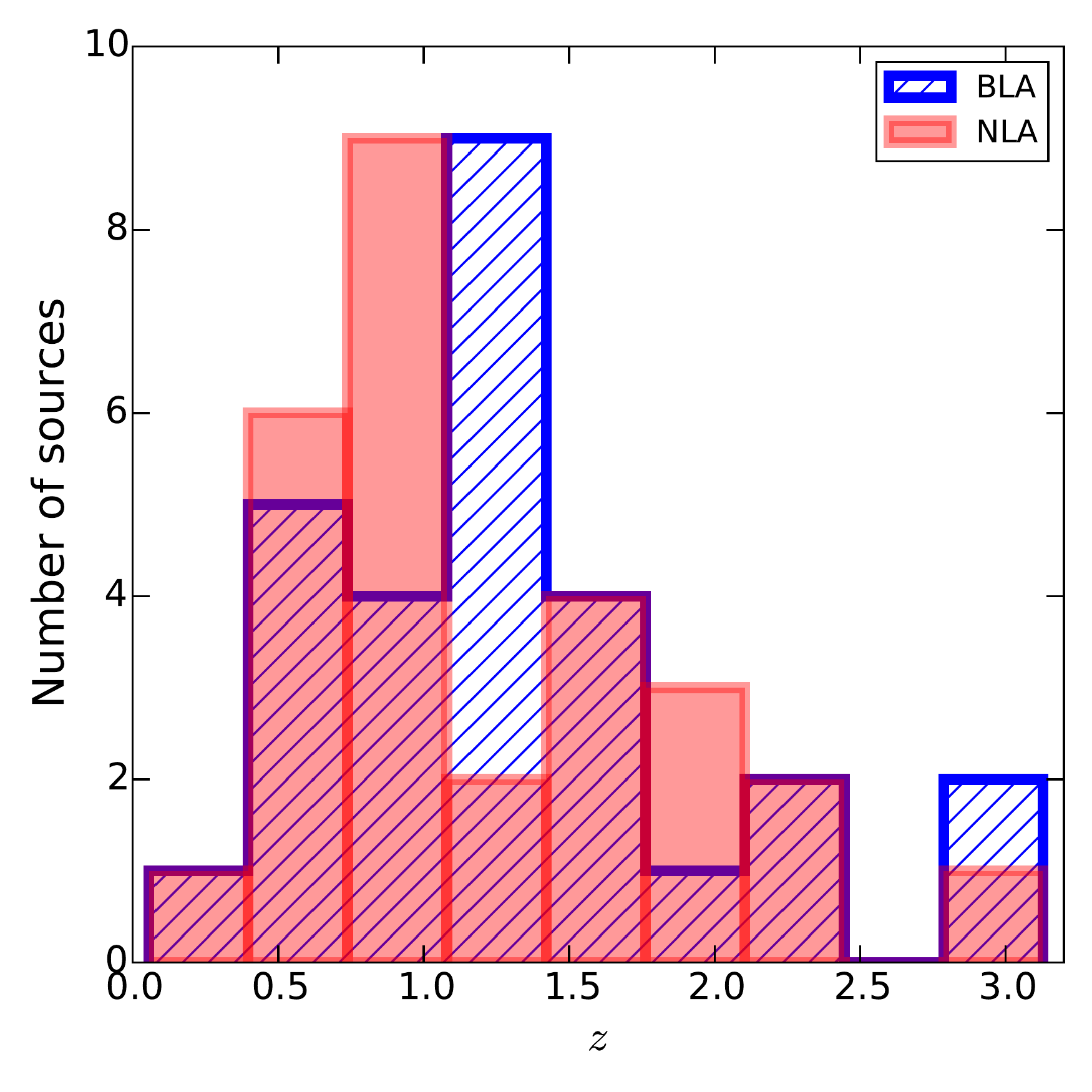}{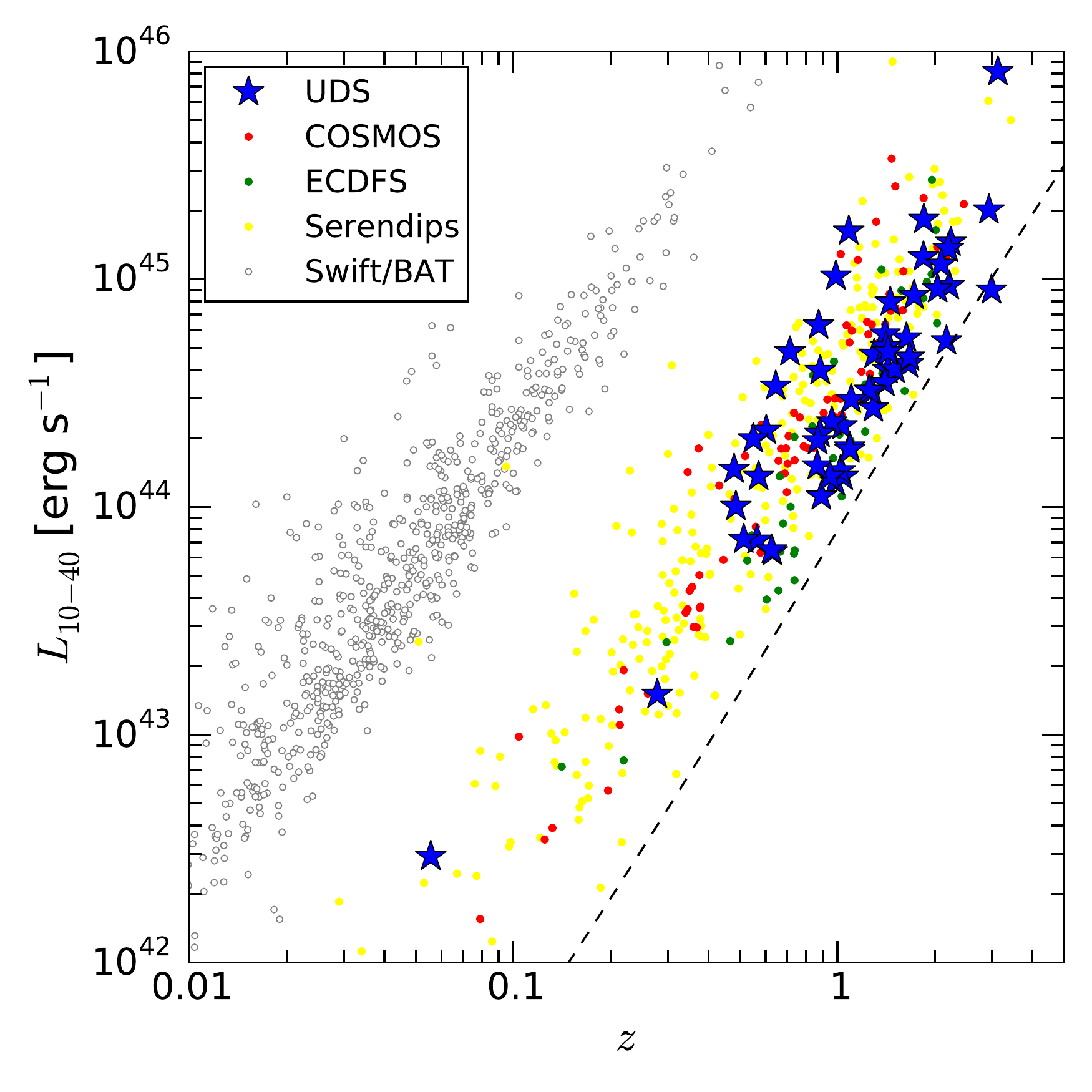}
\caption{\textbf{Left.} Redshift distribution for the 56 sources with an optical counterpart, divided into broad line AGN (blue hatched histogram) and narrow line AGN (red histogram). \textbf{Right.} $10-40$ keV luminosity (not corrected for absorption, although we expect the role of obscuration to be negligible at these energies) as a function of redshift for the NuSTAR surveys. Blue stars are the UDS97 sources. The NuSTAR COSMOS sample \citep{civano15} is indicated by red dots, green dots indicate the NuSTAR ECDFS catalog \citep{mullaney15}, yellow dots the 40-month NuSTAR Serendipitous sample \citep{lansbury17} and gray dots the 70-month \textit{Swift}/BAT catalog \citep{baumgartner13}. The dashed line is the flux limit of the UDS survey at 50\% of completeness. \label{fig:z}}
\end{figure*}

\cite{akiyama15} provide optical counterparts for a large fraction of the SXDS catalog of \cite{ueda08}, and an optical spectrum is also available for the counterpart of uds7, from the BOSS survey \citep{dawson13}. We have redshifts for 56 sources (84\%), of which 48 are spectroscopic and 8 are photometric. We split these 56 redshifts into broad line AGN (BLAGN) and narrow line AGN (NLAGN). The category is directly defined in the \cite{akiyama15} catalog for spectroscopic redshifts based on a FWHM threshold of 1000 km s$^{-1}$, while for photometric redshifts only the ``QSO'' or ``GAL'' templates are specified. We consider then objects best-fitted by a ``QSO'' template as BLAGN, and objects best-fitted by a ``GAL'' template as NLAGN. Out of 56 redshifts, we have 28 BLAGN and 28 NLAGN. The median redshift of our sample is $\langle z\rangle = 1.092$, while $\langle z_{\rm BLAGN}\rangle = 1.272$ and $\langle z_{\rm NLAGN}\rangle = 1.003$. In Figure \ref{fig:z} (left panel) we show the redshift distribution of the sample, while in Figure \ref{fig:z}, (right panel) we show how our sources compare with other NuSTAR Extragalactic Surveys like the COSMOS, ECDFS and Serendipitous ones in the $L_{10-40}-z$ plane. This luminosity is computed from the F band flux without correcting for absorption. We notice that, while the NuSTAR Serendipitous survey \citep{lansbury17} reaches slightly higher redshifts, we detect the highest redshift source among tiered NuSTAR deep surveys (uds67 at $z_{\rm spec} = 3.128$). This may be due to the chosen reliability threshold, which allows fainter sources to be detected. Moreover, the fact that only one source with an optical counterpart is detected at or below the 50\% of completeness is likely due to the chosen flux cut in matching the NuSTAR sources with the SXDS catalog. This flux cut is a factor of three lower than the flux limit at the 50\% of completeness.

\section{Obscuration properties of the NuSTAR UDS97 sample}
We exploit the available redshift information performing a broadband ($0.5-24$ keV) X-ray spectral analysis of all 56 sources in our UDS97 catalog with an optical counterpart.

\subsection{Extraction of X-ray spectra}
We extract FPMA and FPMB spectra with the NuSTARDAS task \texttt{nuproducts}, while NuSTAR background spectra are produced using the \texttt{nuskybgd} software. Following the methodology of Zappacosta et al. (submitted), for any source and pointing in which it is found, we extract the number of counts, background counts and average exposure in the F band for a range of extraction radii. This allows us to get a signal-to-noise ratio (SNR) profile as a function of extraction radius for each pointing and each FPM; the single SNR(r) profiles are then averaged together, weighted with the exposure time at the position of the source in every observation. Finally, the extraction radius for FPMA and FPMB is chosen as the radius where both the weighted SNR and the net counts profiles are approximately peaking. Single data products are summed together using standard tools like \texttt{mathpha}, \texttt{addarf} and \texttt{addrmf}, where the ARFs and RMFs are weighted using the fraction of total counts their respective observation is contributing. 
\par \chandra\ observations of the UDS field are downloaded from the public archive and reduced through the standard pipeline, using the \texttt{chandra\_repro}, \texttt{specextract} and \texttt{combine\_spectra} tasks within the CIAO software (version 4.9, CALDB version 4.7.3). Circular extraction regions with a radius of 2$\arcsec$ are used, while we employ annuli centered on the source position with an internal radius of 3$\arcsec$ and external radius of 10$\arcsec$ to extract background spectra.
\par The \xmm\ observations of the UDS field are downloaded and reduced using the Science Analysis Subsystem (SAS; version 16.0.0) tasks \texttt{epproc/emproc} and filtering every event file for high background time intervals. PN data are always preferred to MOS data when available (i.e., when the source is not falling on a PN gap), while MOS1 and MOS2 spectra are summed together. We use the SAS \texttt{evselect} task to define optimized extraction radii for sources, while background spectra are extracted from nearby circular regions on the same chips as the sources. Finally, we use the SAS task \texttt{epicspeccombine} to produce summed sources and background spectra, ancillary and response files.  Final spectral products are grouped to a minimum of 3 counts/bin with the \texttt{grppha} tool for each telescope.

\subsection{Broadband spectral fitting}\label{sec:obsfrac}
We fit the NuSTAR spectra jointly with the \xmm\ and, when available, \chandra\ ones, adopting the Cash statistic \citep{cash79}. The adopted spectral model is composed of a primary power-law emission with a fixed photon index $\Gamma = 1.8$, taking into account the possible photoelectric absorption and Compton scattering (i.e., a \texttt{plcabs} model), plus allowing for up to a few percent of the primary power law to be scattered into the line of sight. The whole nuclear emission is then absorbed by a Galactic column density  \citep[$N^{\rm gal}_{\rm H} = 2.08 \times 10^{20}$ cm$^{-2}$,][]{kalberla05}, and cross-calibration between instruments is accounted for using a multiplicative factor, and solving for it for each source. In XSPEC notation, our baseline model is given by \texttt{const*pha*(plcabs+const*zpow)}.
\begin{figure*}
\centering
\includegraphics[width=0.3\textwidth]{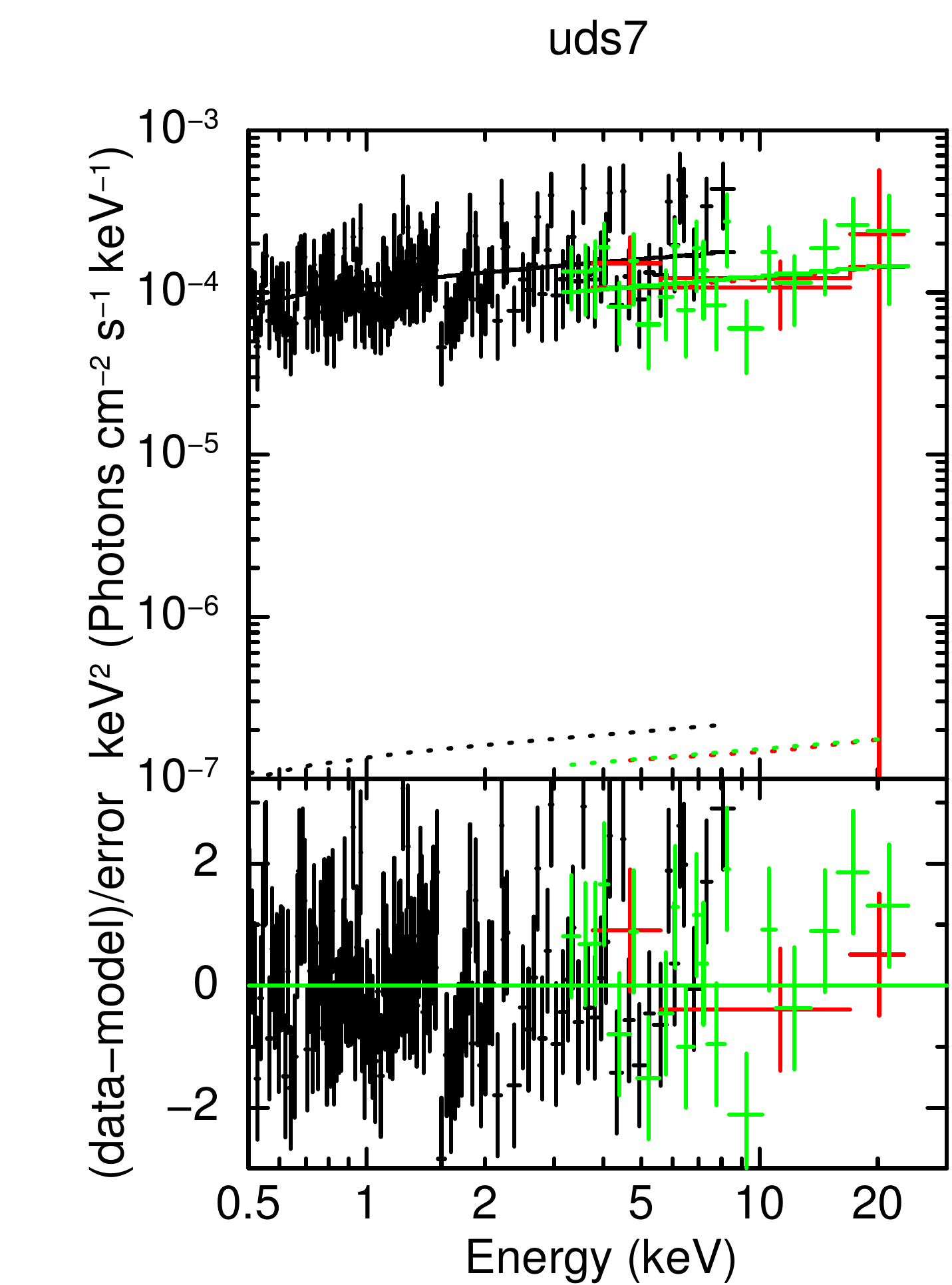}
\includegraphics[width=0.3\textwidth]{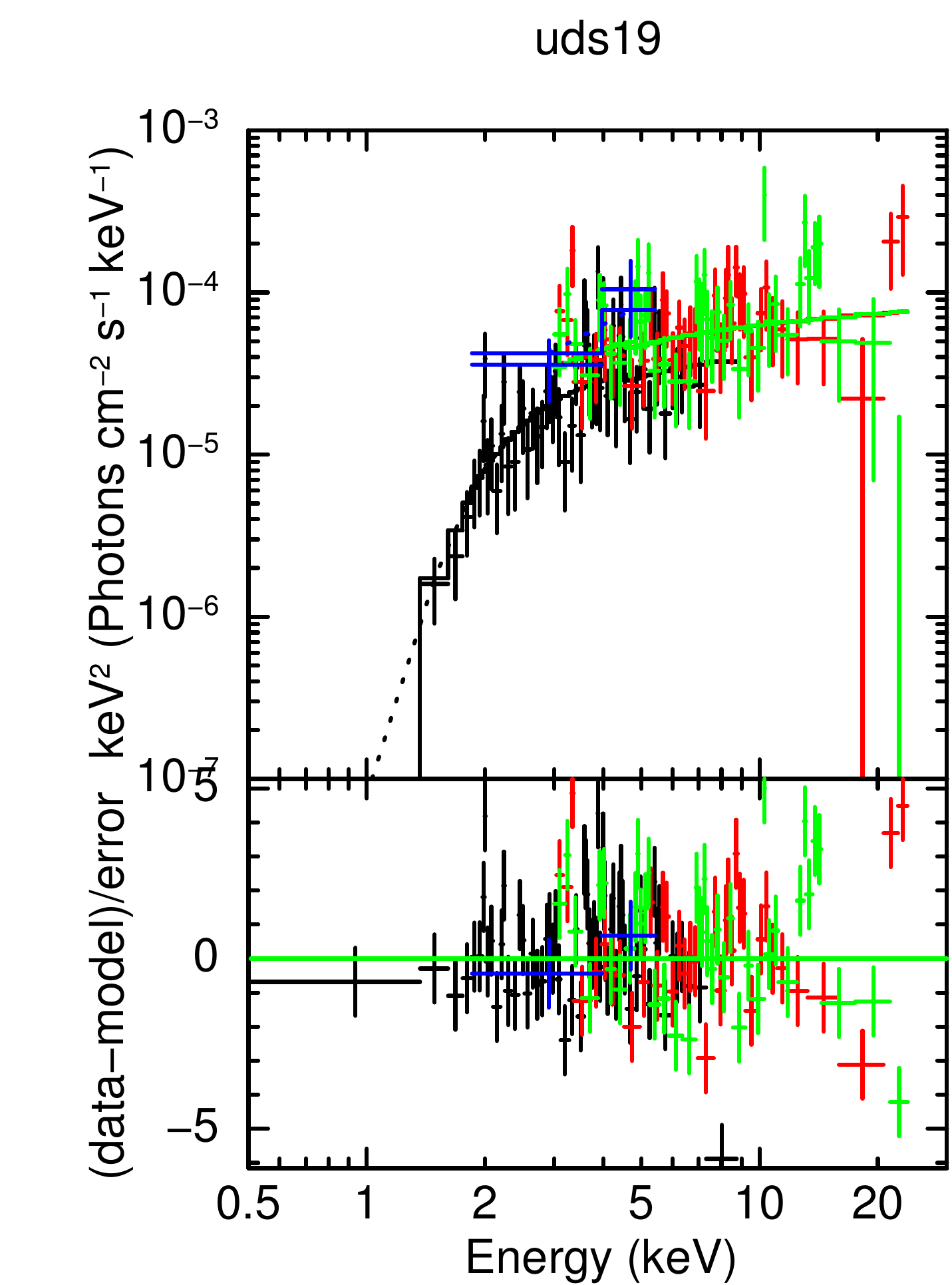}
\includegraphics[width=0.3\textwidth]{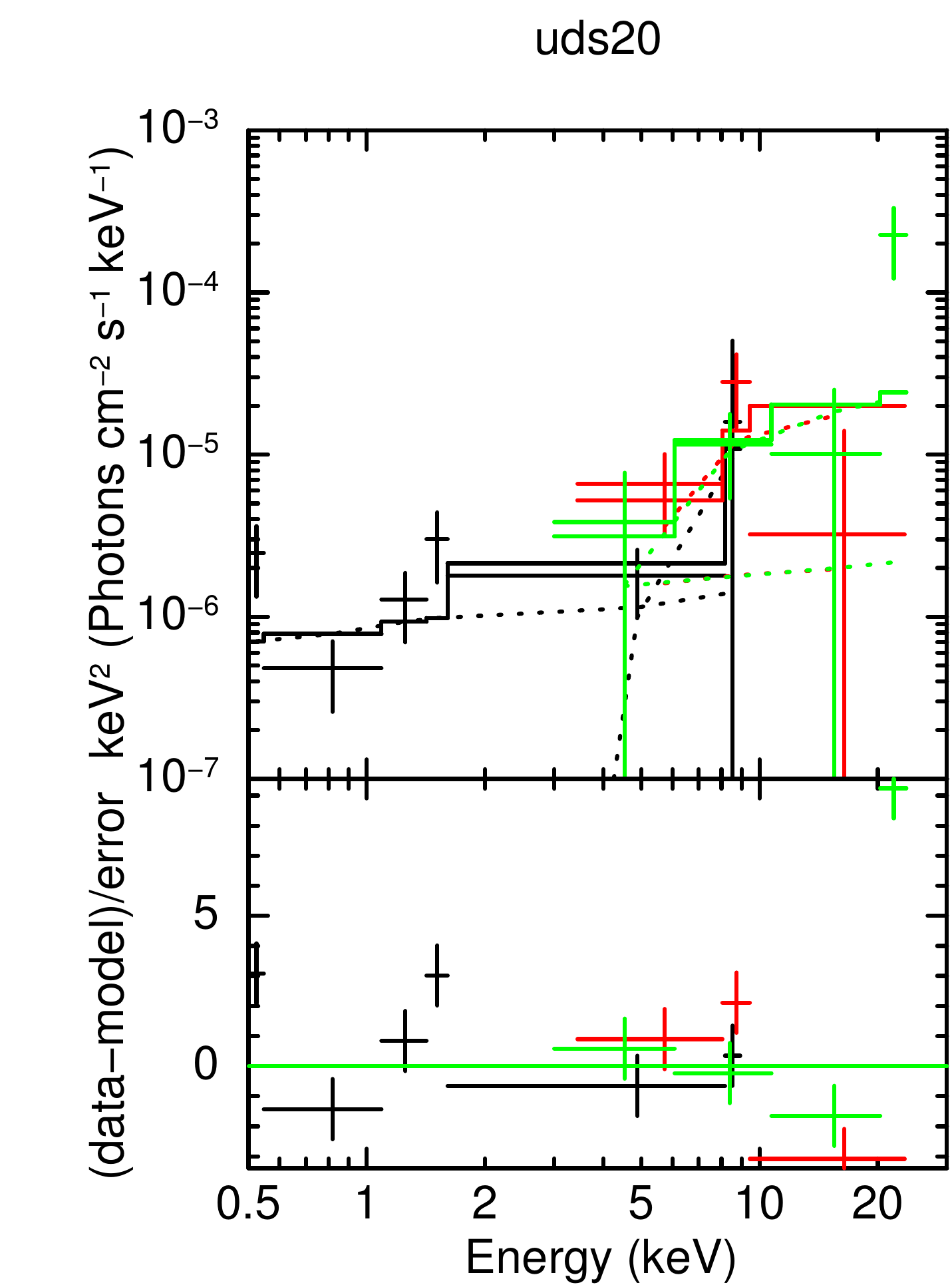}
\caption{Three examples of spectra from the NuSTAR UDS97 sample, showing different levels of obscuration. From left to right, an unobscured, a mildly obscured ($N_{\rm H} \sim 10^{23}$ cm$^{-2}$), and a CT AGN are shown. Each panel shows the  $0.5-24$ keV $\nu F\nu$ spectrum unfolded with the best-fit model (the different components of the model are labeled with dotted lines), and its residuals. Black points are \xmm\ data, while red and green are from NuSTAR FPMA and FPMB, respectively. Where also \chandra\ data is available (middle panel), it is labeled in blue. \label{fig:ex_spectra}}
\end{figure*}
An example of typical unobscured ($N_{\rm H} < 10^{22}$ cm$^{-2}$), mildly obscured ($10^{22} < N_{\rm H} < 10^{24}$ cm$^{-2}$) and heavily obscured spectra ($N_{\rm H} > 10^{24}$ cm$^{-2}$) are shown in Figure \ref{fig:ex_spectra}. A good way to estimate the goodness of fit when using the Cash statistic is running simulations. This is well implemented in XSPEC through the \texttt{goodness} command, which runs a set of $N$ simulated spectra drawn from the best fitting parameters, and fits them again. For each faked spectrum, the comparison with the fitting model is done with a Kolmogorov-Smirnov (KS) test, and the final distribution of KS values is compared with the observed KS. The goodness $G$ is defined as the fraction of simulations resulting in a KS value less than the one observed. The lower the $G$ value, the better the fit. Here, we run 1000 simulations for each source. The resulting distribution of $G$ values along with its cumulative distribution are shown in the left panel of Figure \ref{fig:goodness}, as the yellow histogram and line. As it can be seen in the figure, the distribution peaks at low $G$ values, and $\sim 50\%$ of the sources have a $G$ value less than 20, meaning that 80\% of their simulated spectra had a worse fit with respect to the one performed on real data. However, some sources have $G > 80$. In particular, some of them are bright, unobscured AGN for which the baseline model is not accurate enough to properly model the soft excess. We then add to these sources an \texttt{apec} component, in order to model the residuals seen in their spectra at $E< 1$ keV, and present the aggregated results as the dark blue histogram and cumulative. In this way, $<10\%$ of the sources have $G>80$. The tail of the distribution towards high $G$ values is due to poor counting statistics and cross-calibration uncertainties.
\begin{figure*}
\plottwo{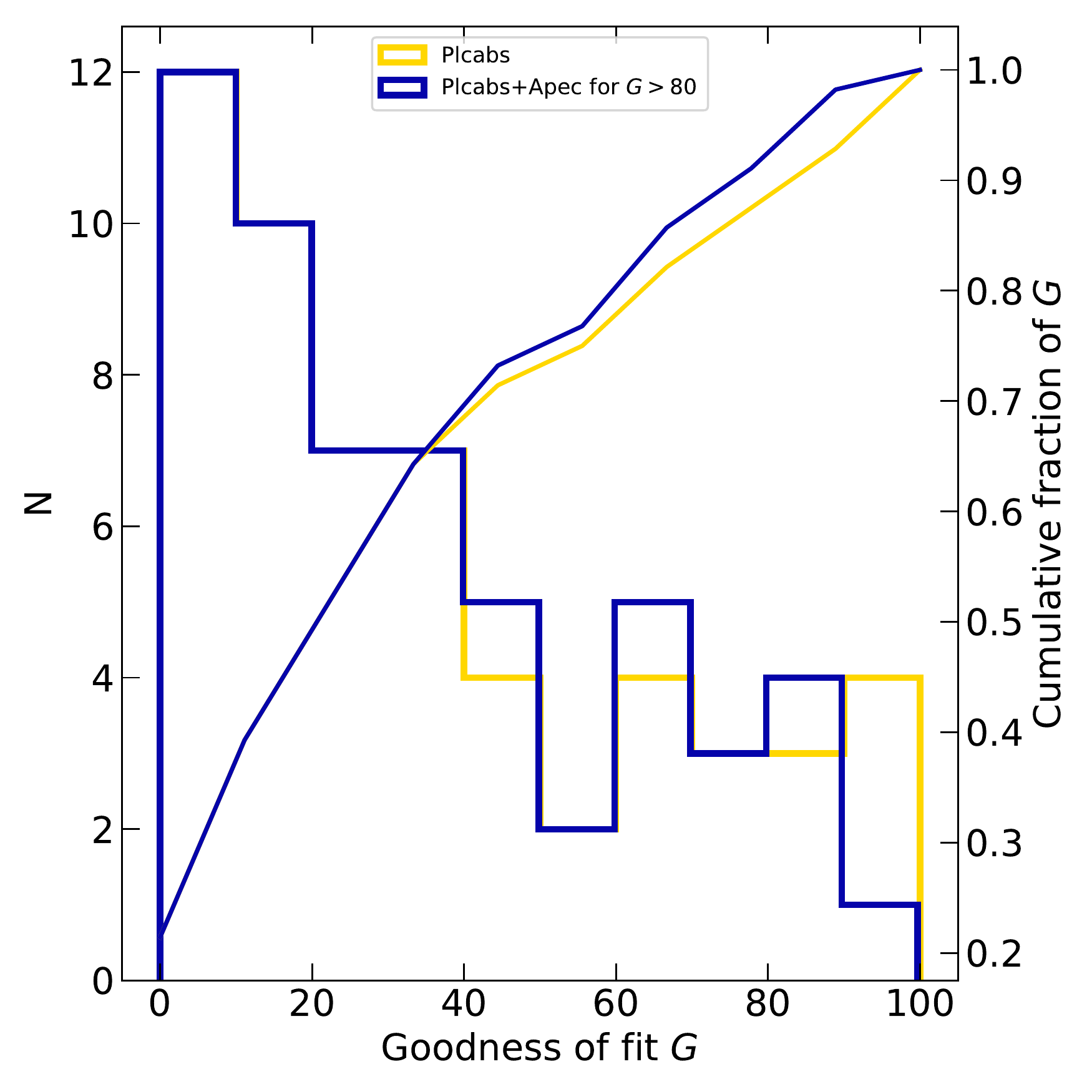}{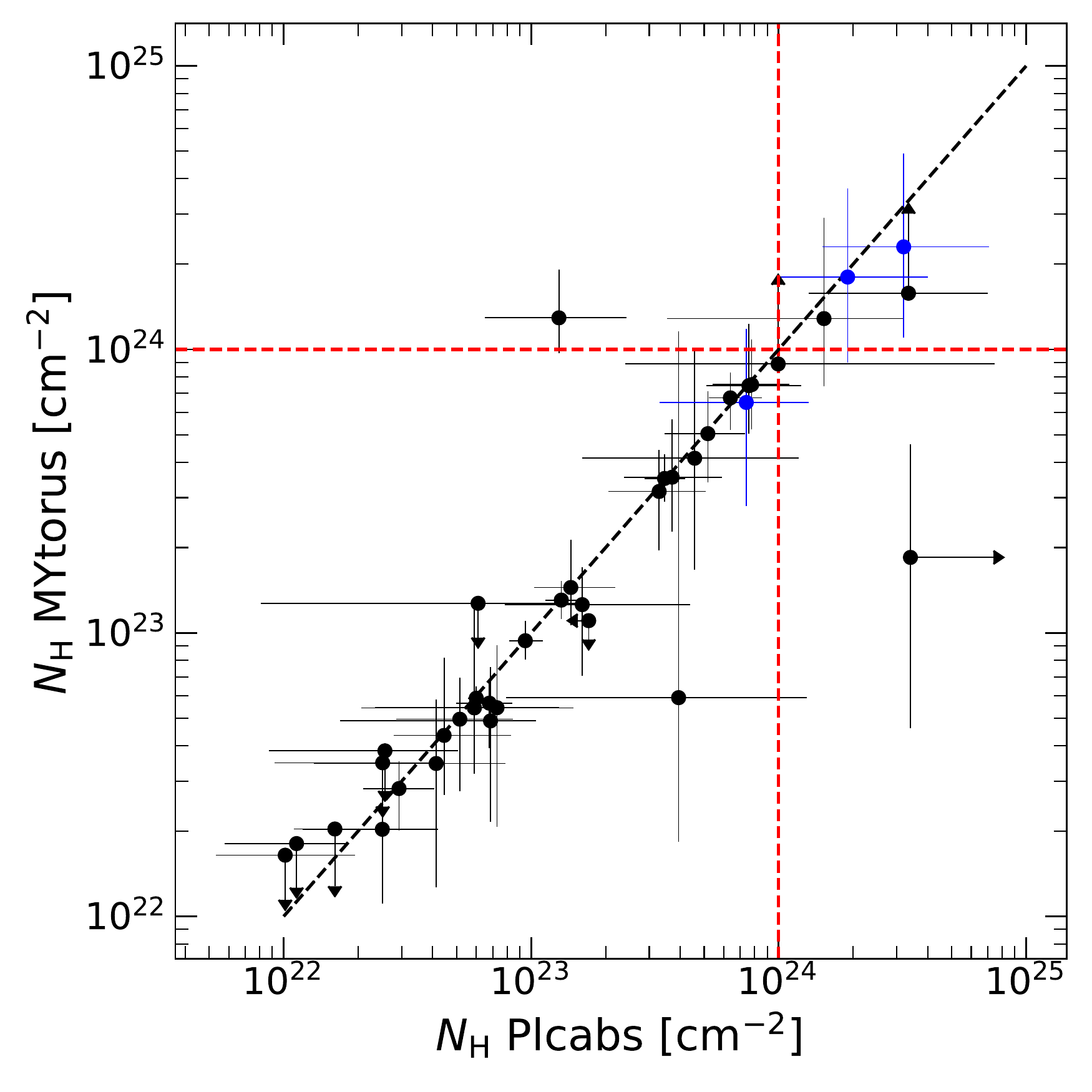}
\caption{\textbf{Left.} Goodness parameter $G$ distribution for the 56 sources with an optical counterpart, fitted with the baseline model (yellow histogram) and with a baseline model plus an \texttt{apec} component for those sources with $G > 80$ (dark blue histogram). Their cumulatives are shown as the yellow and dark blue lines, respectively. About 50\% of the sources have that 80\% of the simulations run in XSPEC resulted in a worse fit. \textbf{Right.} Column density as measured with the baseline model and the MYTorus model. Uncertainties are at the 90\% confidence level. The dashed black line is not a fit to the data, and marks the 1:1 relation. The dashed red lines mark the CT threshold for both models. Sources marked with blue points are obscured added to the sample with the HR analysis, for which some assumptions on the redshift are needed (see \S\ref{sec:uds59} and \S\ref{sec:spec}). \label{fig:goodness}}
\end{figure*}
\par We further check the robustness of the $N_{\rm H}$ values derived with the baseline model, adopting a more realistic model like MYTorus \citep{murphy09}. This model is more appropriate than \texttt{plcabs} when dealing with high column densities ($N_{\rm H} \sim 10^{24}$ cm$^{-2}$), since it self-consistently takes into account Compton scattering and line fluorescence. In this second broadband spectral analysis, we employ a ``default'' MYTorus model, with the line of sight angle set to 90 degrees, plus a scattered power-law, similarly to what was done with the baseline model. In XSPEC notation, the model is \texttt{const*pha*(zpow*MYTZ+MYTS+MYTL+const*zpow)}, where the different components of the MYTorus model stand for the absorption, scattering and line fluorescence ones, respectively. Unfortunately, a direct comparison of the column densities derived from the baseline model and from MYTorus is possible only for those sources having $N_{\rm H} > 10^{22}$ cm$^{-2}$, since MYTorus is designed to deal with CT absorbers and does not allow $N_{\rm H}$ to be lower than $10^{22}$ cm$^{-2}$. A comparison of the column densities derived by the two models is presented in the right panel of Figure \ref{fig:goodness}. An excellent concordance between the two models is seen, except for two sources, uds46 and uds66, which are significantly displaced from the 1:1 relation marked by the black, dashed line.
\par Looking at the fitting of these two sources in detail, we find that for uds46 the baseline model allows a CT solution which is perfectly consistent with the one found with the MYTorus model and is the absolute minimum of the Cash parameter space, while the Compton-thin solution is a relatively deep local minimum. Regarding uds66, the baseline model returns, together with an extremely high column density, also a suspiciously high cross-calibration constant FPMA/EPIC-PN $\sim 22$, which encapsulates both the possible source variability and the cross-calibration uncertainties. When fitted with MYTorus, this solution is disfavored against a Compton-thin solution ($N_{\rm H} \sim 2 \times 10^{23}$ cm$^{-2}$), with a more reasonable cross-calibration constant of FPMA/EPIC-PN $= 2.2^{+4.5}_{-1.3}$. In addition to this, a Compton-thin solution (with a lower CSTAT value with respect to the CT solution with the same model) can be found also with the baseline model. We also verified that this source is the only one returning such an unphysical value for the multiplicative constant. In both cases, the results from MYTorus can be reproduced with the baseline model by applying some fine tuning, and these results provide a better fit to the data.

\subsection{Adding the Hardness Ratio information}\label{sec:spec}

Additional constraints on the obscuration properties of our sample come from the analysis of the hardness ratio (HR), defined as
\begin{equation}
\rm HR= \frac{\rm  H - S}{\rm H + S}.
\end{equation}

We then calculate the HR for the whole UDS97 sample using the Bayesian Estimator for Hardness Ratios \citep[BEHR,][]{park06}, which uses the counts in the S and H NuSTAR bands, and compares the results with the ones coming from the broadband spectral analysis.
At fixed $N_{\rm H}$, the HR changes with redshift, and as such is not possible to infer a unique estimate on $N_{\rm H}$ without a redshift. Nonetheless, sources which show a very hard spectrum (HR $\sim 1$) are good candidates to be highly obscured objects, independently of their redshift \citep{lansbury14,lansbury15}. 
\begin{figure*}
\plottwo{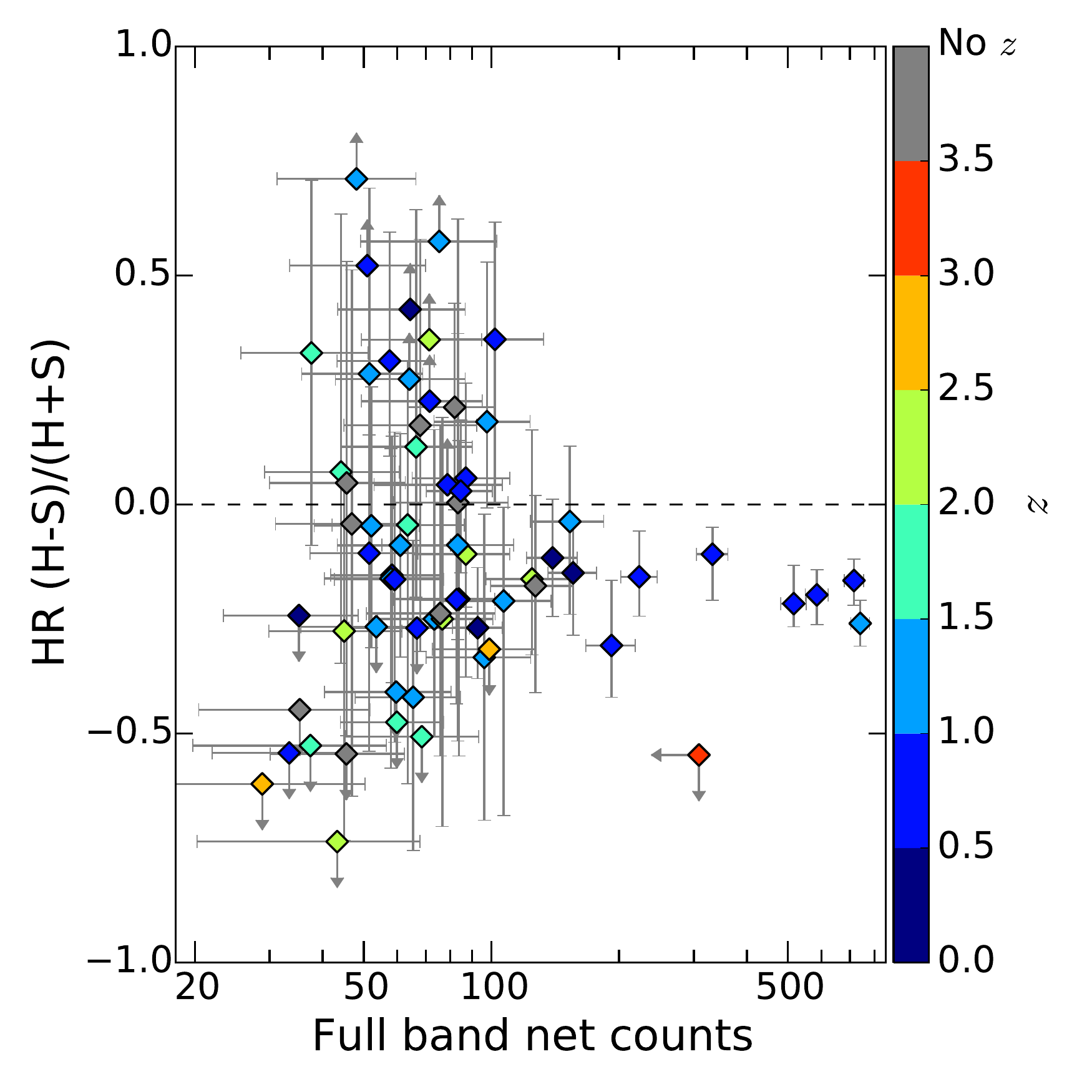}{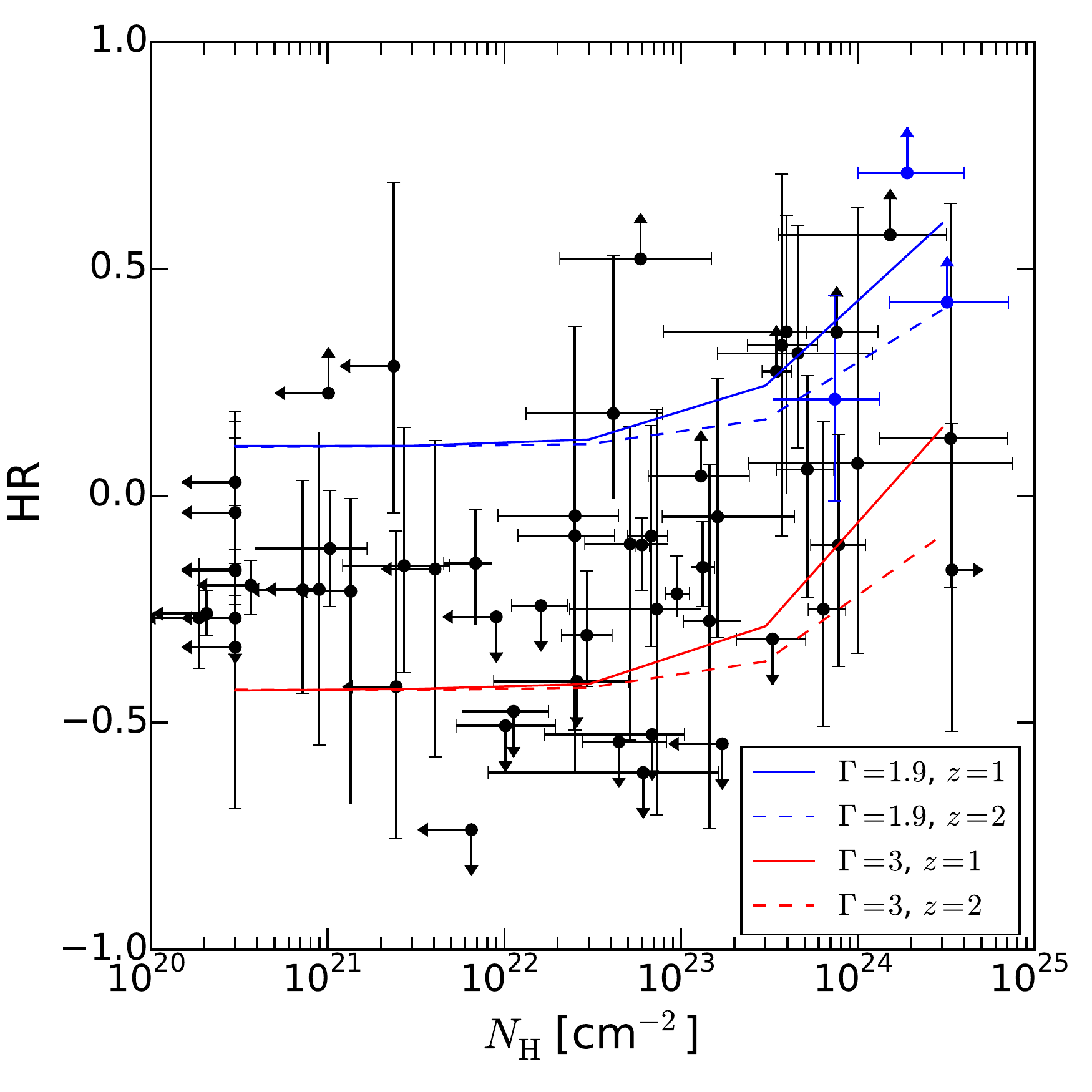}
\caption{\textbf{Left.}  NuSTAR hardness ratio of the whole UDS97 sample as a function of full band ($3-24$ keV) net counts. Points are color-coded with their redshift, where gray points are those without a secure redshift association. A tail of very bright sources at $z<1$ is seen with soft HR, while no clear trend is evident at low counts. The dashed horizontal line marks HR$=0$. \textbf{Right.} Hardness Ratio as a function of obscuring column density, as measured by our baseline model with \texttt{plcabs} (uncertainties at $1\sigma$) for the UDS97 sample, compared with the trend expected from a \texttt{plcabs} model with $\Gamma=1.9$ (blue lines) or an unrealistic $\Gamma=3$ (red lines), at both $z=1$ (solid) and $z=2$ (dashed). The models predict, at a fixed $\Gamma$, the relation to be flat till $N_{\rm H} \sim 10^{23}$ cm$^{-2}$, and an increasing HR for increasing $N_{\rm H}$ at higher column densities. The two blue points mark the $N_{\rm H}$ measured for uds47 and uds63, adopting a \texttt{plcabs} model. \label{fig:nh_hr}}
\end{figure*} 
\par We first test if the low number of net counts could bias our HR measurements. In the left panel of Figure~\ref{fig:nh_hr} the HR is plotted against the NuSTAR full band net counts. While a tail of very bright sources seems to show soft HR, there is no clear trend between the number of counts and the spectral shape, indicating that the HR analysis is not biased towards, or against, a particular level of obscuration when dealing with very few ($\sim$ tens) counts. 
\par Secondly, we test if the HR is effectively tracing the obscuration of our sample, plotting the HR coming from BEHR as a function of the column density $N_{\rm H}$ as measured with our baseline model. As it can be seen from the right panel of Figure \ref{fig:nh_hr}, there is a qualitative concordance between our sample and the trend expected from models, since higher column densities are generally measured for objects with higher HR; moreover, the HR becomes sensitive to a column density change only for $N_{\rm H} > 10^{23}$ cm$^{-2}$.
\par Eight sources in our sample have a lower limit (at $1\sigma$ confidence level) on the HR: uds13, uds30, uds42, uds46, uds47, uds48, uds58, and uds63. We note that uds47 and uds63 lack robust redshift associations, and as such have not been included in the broadband spectral analysis of Section \S \ref{sec:obsfrac}, but have a best-fit HR $= 1$, indicating high obscuration.
\par In particular, the \chandra\ spectrum of uds47 shows a prominent line feature. Assuming that the line is due to the neutral iron K$\alpha$ transition, a redshift of $z = 0.45^{+0.17}_{-0.11}$ is obtained with the MYTorus model. Applying an \textit{ad hoc} model like \texttt{zwabs*zpow+zgauss}, the redshift can be constrained to be $z = 0.47^{+0.06}_{-0.08}$. Assuming this redshift, the spectrum of uds47 is then fitted with the baseline and toroidal models, being CT according to both.
\par Like uds47, uds63 does not have a secure redshift, in this case because it is associated with three \xmm\ sources within 30$\arcsec$. The farthest one is also the faintest at $4.5-10$ keV, while the other two are formally a blend, as described in the text (see \S\ref{sec:match_xmm}), lying at $z_1 = 1.5$ (photometric) and $z_2 = 0.568$ (spectroscopic). Since we are not able to decide which of the two is contributing most to the NuSTAR flux, we extracted both spectra and fitted them separately with the NuSTAR data. In both cases the fit returned a Compton-thick column density, and in this analysis we chose to fit the counterpart closest to the NuSTAR centroid, which provides also the best fit, adopting the redshift $z_1 = 1.5$. Assuming this photometric redshift, the source is CT according to both the baseline and MYTorus models. These two sources, together with uds59 (see \S\ref{sec:uds59}), are marked in the right panel of Figure \ref{fig:goodness} and in the right panel of Figure \ref{fig:nh_hr} as the blue points.
\par
\begin{figure}
\plotone{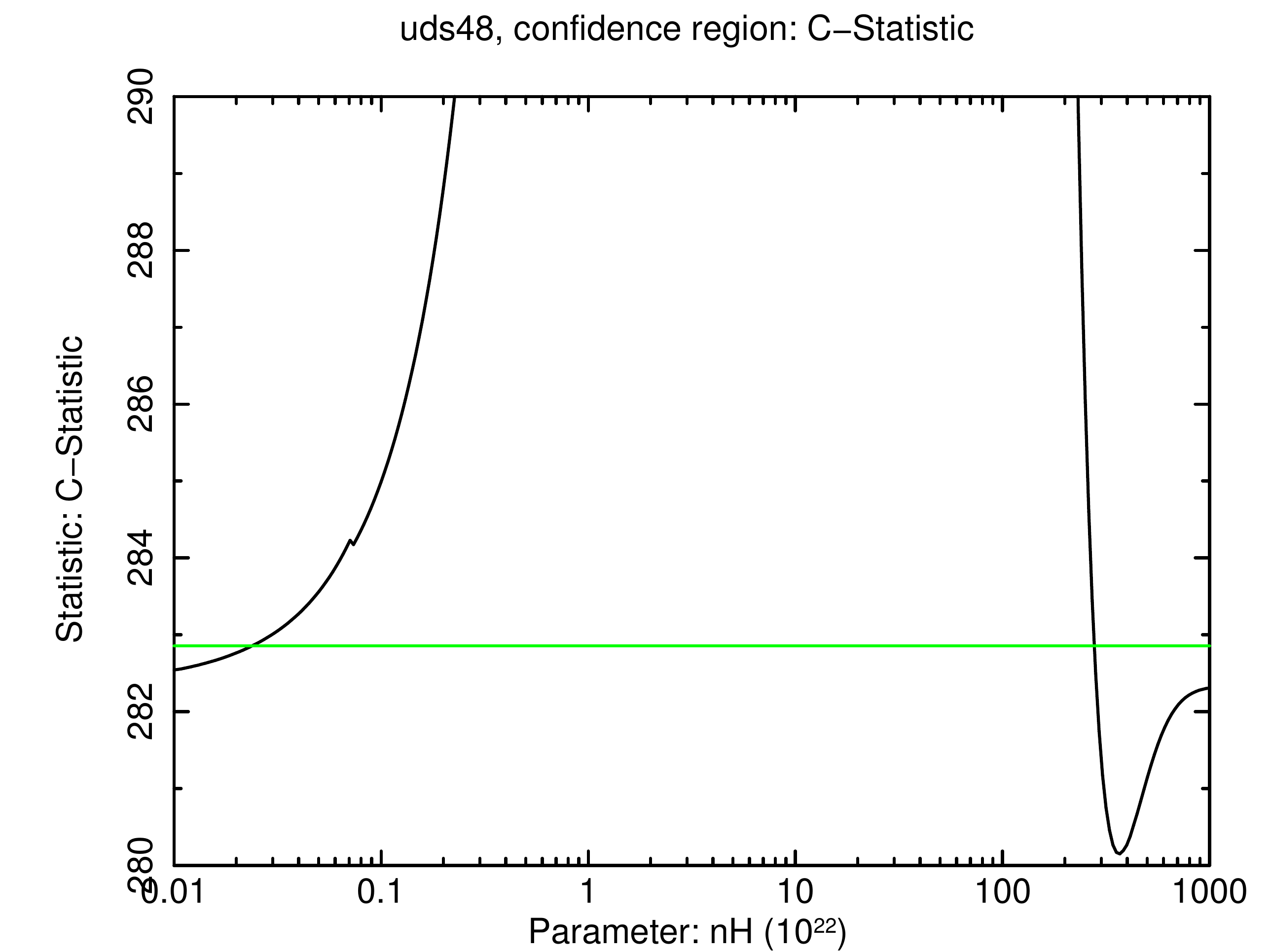}
\caption{C-Statistic parameter as a function of the column density for uds48, fitting NuSTAR + \xmm\ + \chandra\ data. The green horizontal line is the limiting CSTAT for the 90\% confidence level uncertainty. Statistically acceptable solutions are those with a CSTAT parameter below the green line. The model adopted is \texttt{plcabs}, because MYTorus does not allow $N_{\rm H} < 10^{22}$ cm$^{-2}$. An unobscured solution is statistically indistinguishable from a CT one; the best fit, however, is obtained with $N_{\rm H} \gtrsim 3.5 \times 10^{24}$ cm$^{-2}$ and we assume this source to be a CT source. \label{fig:ambig}}
\end{figure}

 The majority of the eight lower limits identified by the BEHR have $N_{\rm H} \gtrsim 10^{23}$ cm$^{-2}$ based on the spectral fitting. All of them are present in the right panel of Figure \ref{fig:goodness} but one, uds48, which is found completely unobscured by the baseline model. A closer look at this source reveals that the \chandra\ and \xmm\ spectra are fit by pure power laws without any sign of obscuration aside from the Galactic one, along the line of sight. Two formally indistinguishable scenarios are possible: one in which the source is totally unobscured, and one in which NuSTAR captures the Compton reflection hump while \xmm\ and \chandra\ detect only the scattered power law emission. We show, in this case, how the CSTAT varies as a function of column density parameter in Figure \ref{fig:ambig}. From the figure, the CT solution seems to be preferred by the data, since the two scenarios have the same number of spectral bins, degrees of freedom and free parameters in the fit. From this, it follows that also the Akaike information criterion \citep[AIC,][]{akaike74} prefers the CT solution. We note that this plot has been obtained adopting the baseline model, since MYTorus does not allow $N_{\rm H} < 10^{22}$ cm$^{-2}$.

In summary, adding the HR information allows us to identify two additional sources as candidates CT AGN (uds47 and uds63), while doubts remain on the nature of uds48, with a slight preference for a CT scenario as well.

\begin{figure*}[!t]
\plottwo{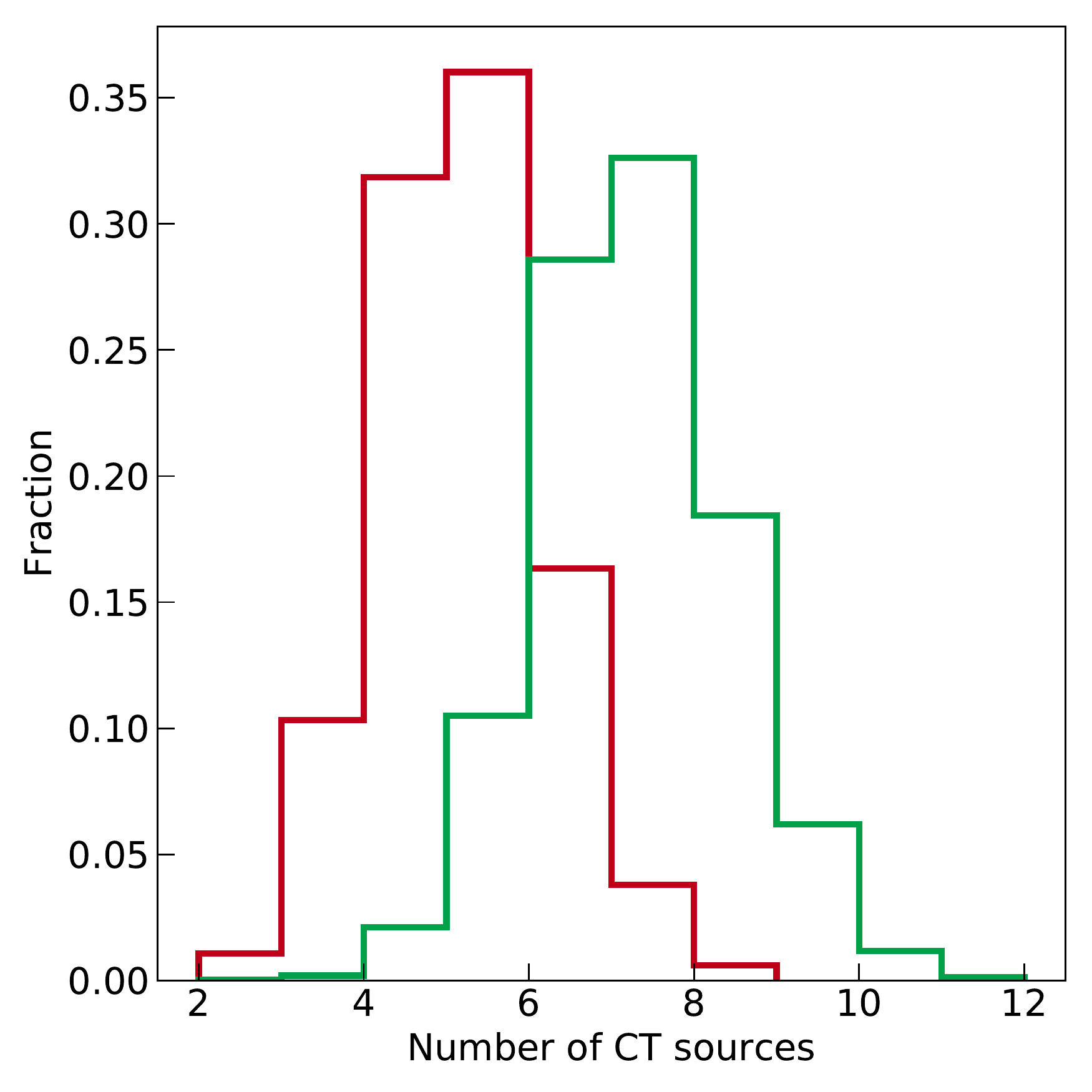}{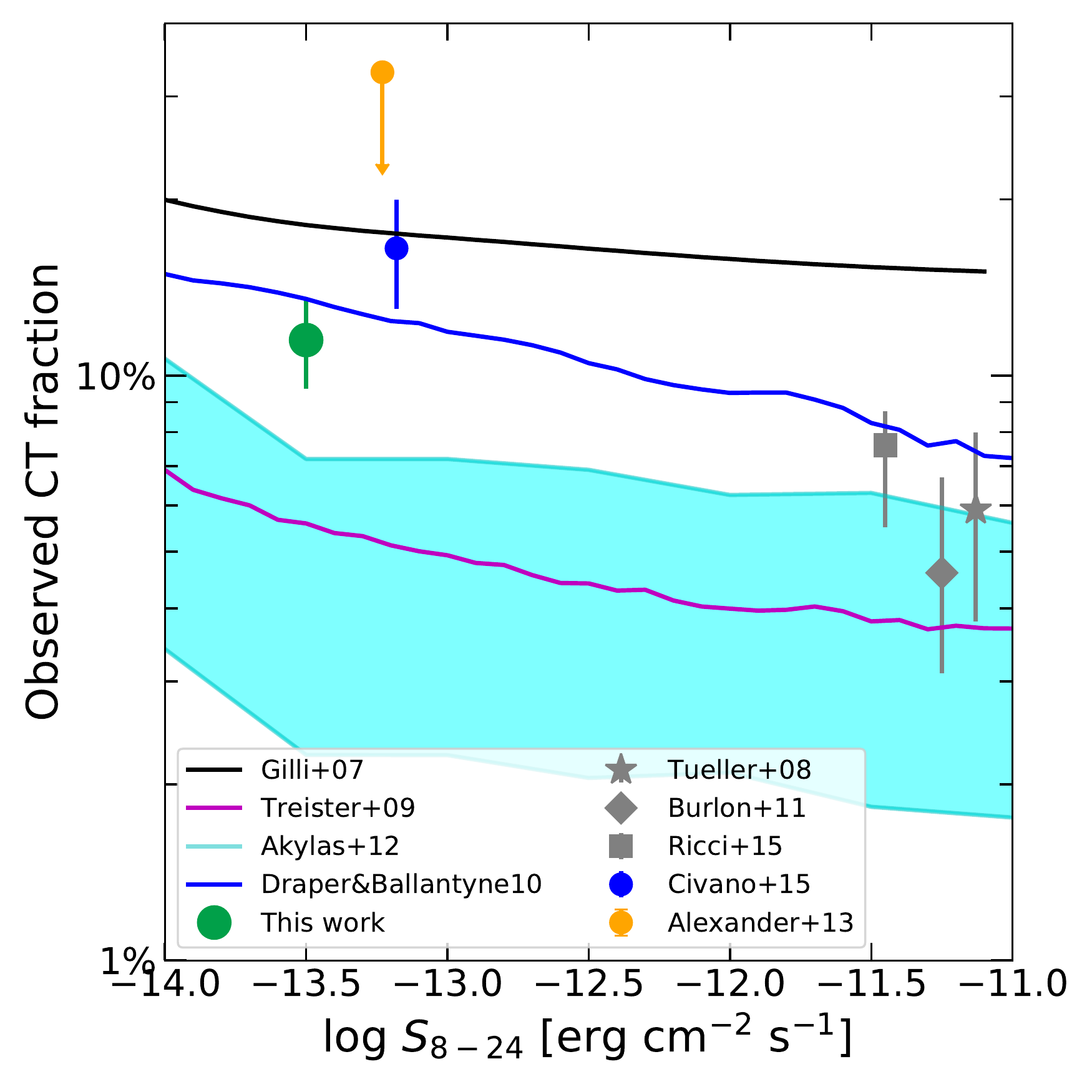}
\caption{\textbf{Left.} Distribution of the number of CT sources for 500 iterations taking into account uncertainties on the single column densities $N_{\rm H}$. The red histogram is obtained considering the UDS97 sample of 56 sources, while the green one is obtained adding to the analysis three heavily obscured sources discussed in the text (uds47, uds59, uds63), for a total sample of 59 sources. \textbf{Right.} Compton-thick fraction as a function of $8-24$~keV flux as measured by NuSTAR in the UDS field in the redshift range $0<z<3$ (green point, which refers to the green histogram in the left panel of this Figure, at a 50\% of completeness flux limit in the $8-24$ keV band). The gray star, diamond and square symbols represent the measurements of \textit{Swift}/BAT by \citet{tueller08}, \citet{burlon11} and \citet{ricci15}, respectively. We note that in \citet{ricci15} the observed CT fraction in the $10-40$ keV band agrees with the \citet{treister09} model, while the $8-24$ keV band adopted here suffers more from absorption, lowering the number of CT sources predicted by the model. The orange point refers to results from the NuSTAR Serendipitous survey \citep{alexander13}, while the blue point is the measurement from \citet{civano15}. Black, blue and magenta solid lines refer to the \citet{gilli07}, \citet{draperballantyne10} and \citet{treister09} models, respectively, while the cyan area refers to the \citet{akylas12} model with a range of intrinsic CT fractions between $15\%-50\%$. \label{fig:ct}}
\end{figure*}

\subsection{The observed CT fraction}
Given the above results, we can now compute the CT fraction of the NuSTAR UDS97 sample taking into account the uncertainties on the column density estimates for each source. We adopt the column densities coming from the baseline model for $N_{\rm H} < 10^{23}$ cm$^{-2}$, and those coming from MYTorus for more obscured sources. This implies that we do not consider uds48 as obscured in the following analysis, despite there being indication that it could be a CT candidate as well. 
\par The $N_{\rm H}$ measurement for a source can be constrained, a lower limit, an upper limit, or unconstrained. In the first case, we adopt the uncertainties coming from the spectral analysis and assume a skewed Gaussian distribution for its probability density function (PDF) of the $N_{\rm H}$, taking into account the asymmetry in the upper and lower uncertainties. In the second case, we truncate the Gaussian, peaked on the observed value, at the 90\% of confidence lower limit\footnote{We define the $\sigma$ of the Gaussian to be $\sigma = (N_{\rm H,obs}-N_{\rm H,LL})/2.706$. When a random value for $N_{\rm H}$ is drawn from the distribution, it is rejected if it is below the lower limit. Adopting a flat PDF between the lower limit and the usually adopted limit of $N_{\rm H} = 10^{25}$ cm$^{-2}$ gives the same results on the CT fraction within the uncertainties.}. In the third case, upper limits are all obtained for unobscured sources. Since the fits are not sensitive to low values of $N_{\rm H}$ (i.e., column densities $N_{\rm H} < 10^{20}$ cm$^{2}$ are statistically indistinguishable\footnote{Since the Galactic $N_{\rm H}$ is $N_{\rm H,Gal} = 2.08 \times 10^{20}$ cm$^{-2}$ (see \S \ref{sec:obsfrac}), unobscured sources show column densities which can be orders of magnitude below the Galactic one. As commonly done in many other previous works, all unobscured AGN with $N_{\rm H} < 10^{20}$ cm$^{-2}$ are put at $N_{\rm H} = 10^{20}$ cm$^{-2}$.}), we adopt a uniform probability distribution between $N_{\rm H} = 10^{20}$ cm$^{-2}$ and the 90\% of confidence upper limit, which goes usually up to $N_{\rm H} \sim \text{few} \times 10^{21}$ cm$^{-2}$. In the last case, similar to the previous one, some sources are totally unobscured, and the fit is completely insensitive to the $N_{\rm H}$ parameter in both directions. In this case, we adopt for simplicity a very narrow half-Gaussian peaked at $N_{\rm H} = 10^{20}$ cm$^{-2}$.
\par With these prescriptions, we run a set of 5000 iterations in which, for each source, we draw a $N_{\rm H}$ value following the underlying assumed PDF. For each iteration, a CT fraction is computed based on the number of sources having $N_{\rm H} > 10^{24}$ cm$^{-2}$. The distribution of the number of CT sources in each iteration is shown in the left panel of Figure \ref{fig:ct} as the red histogram. The average number of CT sources is CT $ =4.7$ with a standard deviation $\sigma=1.1$, which translates to a CT fraction of $f_{\rm CT} = (8.4 \pm 2.0)\%$ ($1\sigma$) given the sample size of 56 sources. If we add those obscured sources coming from both the HR analysis (uds47,uds63, see \S\ref{sec:spec}), and the non-detection by \xmm, uds59 (see \S\ref{sec:uds59}), increasing the sample to 59 sources, the green histogram in the left panel of Figure \ref{fig:ct} is obtained, having an average of $6.8$ CT sources with a standard deviation $\sigma=1.2$, which translates into a $f_{\rm CT} = (11.5\pm2.0)\%$ ($1\sigma$). Consistent results are obtained also in the approximation of adopting symmetric errors (i.e., standard Gaussian distributions for the PDFs).

\section{Discussion on the CT fraction}\label{sec:discussion}
Since we have taken into account the uncertainties on the column density for each source, we are not focusing on a well-defined sample of CT sources, but rather on an average number of sources which are more likely to be CT. On average, for the extended UDS97 sample (i.e., the sample of 56 sources plus the three obscured sources for which we derived or assumed a redshift in different ways), a number of roughly six-seven sources are on average found to be CT.
\par Interestingly, the CT fraction would have been dramatically lower (i.e., $\sim [3\pm2]\%$) in the more conservative sample, UDS99, since on average only one or two of the spectroscopically confirmed CT sources would have been detected. On the other hand, the CT fraction is unlikely to be much higher than $\sim11-12\%$, although few sources may have an HR consistent with a CT obscuration and have not been selected by our criteria, mainly because they are blended, or missing an optical counterpart. 
\par The CT fraction obtained in the UDS field is in agreement with the tentative estimate of C15 based on the Hardness Ratio (HR) distribution of the NuSTAR COSMOS sample, which is $13\%-20\%$, with only one source being confirmed CT from the spectral analysis. It is interesting to note that, if focusing on the UDS99 sample with 43 sources (39 with optical counterpart), and exploiting the HR$-z$ plane as a first-order diagnostic tool to get an estimate of the CT fraction, we would get a fraction of $\sim 15\%$ (i.e., six out of 39 sources with an HR consistent with having $N_{\rm H} > 10^{24}$ cm$^{-2}$), although on average only one of our spectroscopically confirmed CT sources would have been detected in the UDS99 sample. This number, even if remarkably consistent with the spectral analysis result, shows how a robust spectral analysis is required to draw firmer conclusions on the CT fraction, while the use of the HR alone would have yielded uncertain conclusions due to the smaller number of sources. \par A comparison of the observed CT fraction measured by NuSTAR in the UDS field with some models (Figure \ref{fig:ct}, right panel) shows a broad agreement of our result with the predictions of different population synthesis models of the CXB  \citep{gilli07,treister09,draperballantyne10,akylas12}. We again emphasize that here we focus on the observed CT fraction, while a careful estimate of the intrinsic CT fraction is beyond the scope of the paper. Together with the results reported in the figure, we note that the UDS and Serendipitous surveys seem to be probing two different, but complementary, regimes of parameter space with their two different samples. Indeed, \citet{lansbury17b} find that the low redshift ($z<0.07$) CT fraction is unexpectedly high ($\sim 30\%$) compared to model predictions, while the UDS sample covers much higher redshifts, since our CT candidates are almost all between $1 < z <2$, and broadly agrees with model predictions (for $0<z<3$). On the other hand, Zappacosta et al. (submitted) present a thorough and homogeneous broadband ($0.5-24$ keV) spectral analysis of 63 NuSTAR-detected sources with $S_{\rm8-24} > 7 \times 10^{-14}$~erg~cm$^{-2}$~s$^{-1}$ and $\langle z \rangle=0.58$, finding a CT fraction between $1\%-8\%$, which is consistent with our result. However, the right panel of Figure \ref{fig:counts} shows that $\sim$ half of our sources reliably detected in the full band have F-band fluxes lower than their H-band cut, implying an even lower H-band flux for our sources (a factor of $7-10$ with respect to the sources selected in Zappacosta et al., submitted). This is also confirmed by the spectral analysis performed. In this respect, we are probing, also in this case, two different, and possibly complementary, redshift and flux ranges.

\section{Conclusions}\label{sec:conclusions}
In this paper, we presented the NuSTAR survey of the UDS field, consisting of 35 observations performed in two separate passes on the field. The total observing time is 1.75 Ms, over an area of 0.58 deg$^2$. The main results can be summarized as follows:
\begin{itemize}
\item{We detected 43 sources above the 99\% reliability threshold (i.e., UDS99), and 67 sources above the 97\% reliability threshold (i.e., UDS97). We have explored, for the first time, the feasibility of a detection in three new bands, splitting the hard $8-24$ keV band into two narrower bands (H1, $8-16$ keV; H2, $16-24$ keV), and exploiting the broad-band capabilities of NuSTAR looking for sources in the very-hard (VH, $35-55$ keV) band. Very few sources are found in the H2 bands and no sources are detected in the VH band. This is in agreement with the simulations performed, which require larger areas to collect more sources and draw firmer results. Applying this analysis to all the NuSTAR Extragalactic Surveys fields seems a natural follow-up of the work presented here, and will be the subject of a future publication. Therefore, the catalog is restricted to the canonical F, S and H bands for homogeneity with the previous NuSTAR Extragalactic Surveys, and we focused on the UDS97 catalog, where the expected spurious fraction is 3\%.}
\item{We identify one NuSTAR source undetected in lower energy data (uds59, discussed in \ref{sec:uds59}), that is likely heavily obscured. A combination of heavy obscuration and low EPIC exposure at its position may be enough to explain its non-detection by \xmm.}
\item{In order to have a precise view of the obscuration properties of our sample, we combined all the available information coming from a broadband spectral analysis and hardness ratio diagnostic to include all the heavily obscured candidates. We then computed an accurate observed CT fraction taking into account the uncertainties on each $N_{\rm H}$ value and running 5000 iterations of the column density distribution. The final CT fraction is $f_{\rm CT}=(11.5\pm2.0)\%$, considering a sample of 59 sources. This fraction is in agreement with findings from other NuSTAR surveys, and in broad agreement with population synthesis models of the CXB.}
\item{If we adopted the more conservative UDS99 sample, the HR$-z$ plane alone diagnostics would yield a CT fraction of $\sim 15\%$. On the other hand, on average only one of our spectroscopically confirmed CT sources would have been detected in the UDS99 sample, dramatically lowering the observed CT fraction. A robust spectral analysis is key to strengthen the results obtained with the HR alone.}
\end{itemize}

\acknowledgments
We thank the anonymous referee for providing helpful comments, which significantly improved the clarity and robustness of the paper. \par
This work was supported under NASA Contract NNG08FD60C, and made use of data from the NuSTAR mission, a project led by the California Institute of Technology, managed by the Jet Propulsion Laboratory, and funded by the National Aeronautics and Space Administration. We thank the NuSTAR Operations, Software, and Calibration teams for support with the execution and analysis of these observations. This research made use of the NuSTAR Data Analysis Software (NuSTARDAS) jointly developed by the ASI Science Data Center (ASDC, Italy) and the California Institute of Technology (USA). \par This research has also made use of data obtained from the \chandra\ Data Archive and the \chandra\ Source Catalog, and software provided by the \chandra\ X-ray Center (CXC), as well as observations obtained with \xmm, an ESA science mission with instruments and contributions directly funded by ESA Member States and NASA, together with data obtained from the 3XMM \xmm\ serendipitous source catalogue compiled by the 10 institutes of the \xmm\ Survey Science Centre selected by ESA. \par
A. M. and A. C. acknowledge support from the ASI/INAF grant I/037/12/0011/13. W.~N.~B. acknowledges Caltech NuSTAR subcontract 44A-1092750. L.Z. acknowledges financial support uder ASI/INAF contract I/037/12/0. G.B.L. acknowledges support from a Herchel Smith Fellowship of the University of Cambridge. E.T. acknowledges support from CONICYT-Chile grants Basal-CATA PFB-06/2007 and FONDECYT Regular 1160999. R.C.H. acknowledges support from NASA through grant number NNX15AP24G.
\vspace{5mm}
\facilities{NuSTAR, \chandra, \xmm}
\software{NuSTARDAS, XSPEC \citep[v 12.9.1][]{arnaud96}, FTOOLS, SExtractor \citep{bertin96}, CIAO \citep{fruscione06}, Astropy \citep{astropy13}, IDL}

\renewcommand{\theHsection}{A\arabic{section}}
\renewcommand*{\theHtable}{\arabic{section}.\arabic{table}} 
\renewcommand*{\theHfigure}{\arabic{section}.\arabic{figure}}

\appendix

\section{Catalog Description}\label{sec:catalog}
The electronic version of the catalog contains the following information. An extract of the first two rows of the catalog is presented in Table \ref{tab:catalog_example}.

\begin{deluxetable}{ll}
\tabletypesize{\scriptsize}
\tablecaption{Details on the catalog content. \label{tab:catalog_description}}
\tablehead{
\colhead{Col number} & \colhead{Description} \\
}
\colnumbers
\startdata
  1 & NuSTAR source name, following the standard IAU convention, with the prefix ``NuSTAR''. \\
  2 & Source ID. \\
  3 & RA of the source, in the J2000 coordinate system. \\
 4 & DEC of the source, in the J2000 coordinate system. \\
  5 &3--24 keV band deblended DET\_ML (0 if undetected). \\
  6 &3--24 keV band vignetting-corrected exposure time at the position of the source.\\
  7 &3--24 keV band total counts in a 20$\arcsec-$radius circular aperture.\\
  8& 3--24 keV band deblended background counts in a 20$\arcsec-$radius circular aperture.\\
  9& 3--24 keV band net counts (3$\sigma$ upper limit if undetected).\\
  10& 3--24 keV band positive count error computed using Gehrels statistic (0 if undetected).\\
  11& 3--24 keV band negative count error computed using Gehrels statistic (0 if undetected).\\
  12& 3--24 keV band count rate in a 20$\arcsec-$radius circular aperture (3$\sigma$ upper limit if undetected).\\
  13& 3--24 keV band aperture-corrected flux (3$\sigma$ upper limit if undetected).\\
  14& 3--24 keV band positive flux error ($-99$ if undetected).\\
  15&3--24 keV band negative flux error ($-99$ if undetected).\\
  16& 3--8 keV band deblended DET\_ML (0 if undetected).\\
  17&3--8 keV band vignetting-corrected exposure time at the position of the source.\\
  18& 3--8 keV band total counts in a 20$\arcsec-$radius circular aperture.\\
  19& 3--8 keV band deblended background counts in a 20$\arcsec-$radius circular aperture.\\
  20& 3--8 keV band net counts (3$\sigma$ upper limit if undetected).\\
  21& 3--8 keV band positive count error computed using Gehrels statistic (0 if undetected).\\
  22& 3--8 keV band negative count error computed using Gehrels statistic (0 if undetected).\\
  23& 3--8 keV band count rate in a 20$\arcsec-$radius circular aperture (3$\sigma$ upper limit if undetected).\\
  24& 3--8 keV band aperture-corrected flux (3$\sigma$ upper limit if undetected).\\
  25& 3--8 keV band positive flux error ($-99$ if undetected).\\
  26&3--8 keV band negative flux error ($-99$ if undetected).\\
  27& 8--24 keV band deblended DET\_ML (0 if undetected).\\
  28& 8--24 keV band vignetting-corrected exposure time at the position of the source.\\
  29& 8--24 keV band total counts in a 20$\arcsec-$radius circular aperture.\\
  30& 8--24 keV band deblended background counts in a 20$\arcsec-$radius circular aperture.\\
  31& 8--24 keV band net counts (3$\sigma$ upper limit if undetected).\\
  32& 8--24 keV band positive count error computed using Gehrels statistic (0 if undetected).\\
  33&8--24 keV band negative count error computed using Gehrels statistic (0 if undetected).\\
  34& 8--24 keV band count rate in a 20$\arcsec-$radius circular aperture (3$\sigma$ upper limit if undetected).\\
  35& 8--24 keV band aperture-corrected flux (3$\sigma$ upper limit if undetected).\\
  36& 8--24 keV band positive flux error ($-99$ if undetected).\\
  37& 8--24 keV band negative flux error ($-99$ if undetected).\\
  38& Hardness ratio computed with BEHR \citep{park06}.\\
  39& Hardness Ratio lower bound.\\
  40&Hardness Ratio upper bound.\\
  41& \xmm\ primary counterpart in 30$\arcsec$ from the SXDS catalog \citep{ueda08} ($-99$ if no counterpart is found).\\
  42& RA of the \xmm\ counterpart, J2000 coordinate system ($-99$ if no counterpart is found).\\
  43& DEC of the \xmm\ counterpart, J2000 coordinate system ($-99$ if no counterpart is found).\\
  44& \xmm\ ultrasoft band ($0.3-0.5$ keV) count rate ($-99$ if no counterpart is found).\\
  45& \xmm\ ultrasoft band ($0.3-0.5$ keV) count rate uncertainty (1$\sigma$) ($-99$ if no counterpart is found).\\
  46& \xmm\ soft band ($0.5-2.0$ keV) count rate ($-99$ if no counterpart is found).\\
  47& \xmm\ soft band ($0.5-2.0$ keV) count rate uncertainty (1$\sigma$) ($-99$ if no counterpart is found).\\
  48& \xmm\ medium band ($2.0-4.5$ keV) count rate ($-99$ if no counterpart is found).\\
  49& \xmm\ medium band ($2.0-4.5$ keV) count rate uncertainty (1$\sigma$) ($-99$ if no counterpart is found).\\
  50& \xmm\ ultrahard band ($4.5-10.0$ keV) count rate ($-99$ if no counterpart is found).\\
  51& \xmm\ ultrahard band ($4.5-10.0$ keV) count rate uncertainty (1$\sigma$) ($-99$ if no counterpart is found).\\
  52& Distance between the NuSTAR source and \xmm\ primary counterpart ($-99$ if no counterpart is found).\\
  53& Number of \xmm\ counterparts found within 30$\arcsec$.\\
  54& \chandra\ primary counterpart in 30$\arcsec$ from the XUDS catalog (Kocevski et al. submitted) ($-99$ if no counterpart is found).\\
  55& RA of the \chandra\ counterpart, J2000 coordinate system ($-99$ if no counterpart is found).\\
  56& DEC of the \chandra\ counterpart, J2000 coordinate system ($-99$ if no counterpart is found).\\
  57& \chandra\ soft band ($0.5-2.0$ keV) flux ($-99$ if no counterpart is found).\\
  58& \chandra\ soft band ($0.5-2.0$ keV) flux positive error ($-99$ if no counterpart is found).\\
  59& \chandra\ soft band ($0.5-2.0$ keV) flux negative error ($-99$ if no counterpart is found).\\
  60& \chandra\ hard band ($2.0-10.0$ keV) flux ($-99$ if no counterpart is found).\\
  61& \chandra\ hard band ($2.0-10.0$ keV) flux positive error ($-99$ if no counterpart is found).\\
  62& \chandra\ hard band ($2.0-10.0$ keV) flux negative error ($-99$ if no counterpart is found).\\
  63& \chandra\ ultrahard band ($5.0-10.0$ keV) flux ($-99$ if no counterpart is found).\\
  64& \chandra\ ultrahard band ($5.0-10.0$ keV) flux positive error ($-99$ if no counterpart is found).\\
   65& \chandra\ ultrahard band ($5.0-10.0$ keV) flux negative error ($-99$ if no counterpart is found).\\
  66& Distance between the NuSTAR source and \chandra\ primary counterpart ($-99$ if no counterpart is found).\\
  67& Number of \chandra\ counterparts found within 30$\arcsec$.\\
  68& RA of optical counterpart, from \citet{akiyama15} ($-99$ if no counterpart is found).\\
  69& DEC of optical counterpart, from \citet{akiyama15} ($-99$ if no counterpart is found).\\
  70& Spec-z of optical counterparts \citep{akiyama15} ($-99$ if no counterpart is found, 9.999 if no spectroscopic redshift is available).\\ 
  71& Spectroscopic classification of optical counterpart ($-99$ if no counterpart is found).\\
  72 & Photo-z of optical counterparts \citep{akiyama15} ($-99$ if no counterpart is found).\\
  73 & Photometric classification of optical counterpart ($-99$ if no counterpart is found).\\
\enddata
\end{deluxetable}

\floattable
\begin{deluxetable*}{l c c c c c c c c c c c}
\tabletypesize{\normalsize}
\tablecaption{An extract from the UDS catalog. \label{tab:catalog_example}}
\tablehead{
\colhead{Name} & \colhead{Number} & \colhead{R.A.} & \colhead{DEC.} & \colhead{DET\_ML$_{\rm F}$} & \colhead{$E_{\rm F}$} & \colhead{$T_{\rm F}$} & \colhead{$B_{\rm F}$} & \colhead{$N_{\rm F}$} & \colhead{$\sigma^{+,N}_{\rm F}$} & \colhead{$\sigma^{-,N}_{\rm F}$} & \colhead{CR$_{\rm F}$} \\
 \colhead{}  & \colhead{} &  \colhead{[deg]} &  \colhead{[deg]} &\colhead{} & \colhead{[ks]} & \colhead{}& \colhead{}& \colhead{}& \colhead{}& \colhead{}& \colhead{[s$^{-1}$]}\\
 \colhead{(1)}  & \colhead{(2)} &  \colhead{(3)} &  \colhead{(4)} &\colhead{(5)} & \colhead{(6)} & \colhead{(7)}& \colhead{(8)}& \colhead{(9)}& \colhead{(10)}& \colhead{(11)}& \colhead{(12)}\\ \noalign{\vskip -0.5cm}
}
\rotate
\startdata
 J340916-050237.5	&1&	34.1547&	$-5.0438$&	8.186&	157.5	 & 175.0&	137.57	&37.43	&19.13&	17.66	&0.000238\\  \noalign{\vskip 0.5mm}
 J341921-051012.4	&2&	34.3226&	$-5.1701$&	5.720&	223.5  &	216.0&	187.18	&28.82&	21.52&	20.06&	0.000129\\ \noalign{\vskip 0.5mm}
  \hline
\noalign{\vskip 1cm}
 \hline\hline       
\noalign{\vskip 1mm} 
Flux$_{\rm F}$  & $\sigma^{+,\rm Flux}_{\rm F}$ & $\sigma^{-,\rm Flux}_{\rm F}$ & DET\_ML$_{\rm S}$ & $E_{\rm S}$ & $T_{\rm S}$ & $B_{\rm S}$ & $N_{\rm S}$ & $\sigma^{+,N}_{\rm S}$ & $\sigma^{-,N}_{\rm S}$ & CR$_{\rm S}$ & Flux$_{\rm S}$ \\ \noalign{\vskip 0.5mm} 
 [cgs] & [cgs] &  [cgs] & & [ks] & & & & & & [s$^{-1}$] & [cgs] \\\noalign{\vskip 0.5mm} 
(13)     & (14)   &(15) & (16) &(17) &(18)&(19)&(20) & (21) & (22) & (23) & (24)  \\
\noalign{\vskip 0.5mm}  
\hline                   
\noalign{\vskip 1mm}  
$3.6\times 10^{-14}$&	$1.8\times10^{-14}$&	$1.7\times10^{-14}$&	12.314&	169.5	& 98.0&	61.6&	36.4&	14.09&	12.60&	0.000215&	$2.3\times10^{-14}$ \\  \noalign{\vskip 0.5mm}  
 $2.0\times10^{-14}$&	$1.5\times10^{-14}$&	$1.4\times10^{-14}$&	13.877&	237.2&	130.0&	84.46&	45.54&	16.09&	14.62&	0.000192&	$2.0\times10^{-14}$\\ \noalign{\vskip 0.5mm}
 \hline
\noalign{\vskip 1cm}
\hline\hline       
\noalign{\vskip 1mm} 
$\sigma^{+,\rm Flux}_{\rm S}$ & $\sigma^{-,\rm Flux}_{\rm S}$ & DET\_ML$_{\rm H}$ & $E_{\rm H}$ & $T_{\rm H}$ & $B_{\rm H}$ & $N_{\rm H}$ & $\sigma^{+,N}_{\rm H}$ & $\sigma^{-,N}_{\rm H}$ & CR$_{\rm H}$ & Flux$_{\rm H}$ & $\sigma^{+,\rm Flux}_{\rm H}$ \\ \noalign{\vskip 0.5mm} 
 [cgs] &  [cgs] & & [ks] & & & & & & [s$^{-1}$] & [cgs] &  [cgs] \\\noalign{\vskip 0.5mm} 
(25)& (26)&(27)&(28)&(29)&(30)&(31)&(32)&(33)&(34)&(35)&(36)\\
\noalign{\vskip 0.5mm}  
\hline                   
\noalign{\vskip 1mm}  
$8.8\times10^{-15}$& $7.9\times10^{-15}$&	0&	146.4&	83.0&	76.16&	107.85&	0&	0&	0.000737&	$1.6\times10^{-13}$&	$-99$ \\  \noalign{\vskip 0.5mm}  
 $7.2\times10^{-15}$&	$6.5\times10^{-15}$&	0&	212.8	 & 107.0&	103.18&	138.18&	0&	0&	0.000649&	$1.4\times10^{-13}$&	$-99$\\ \noalign{\vskip 0.5mm}
\enddata
\end{deluxetable*}

\bibliographystyle{aasjournal} 
\bibliography{/media/alberto/TOSHIBA2/amasini} 

\listofchanges

\end{document}